\documentclass[aps,reprint,longbibliography,superscriptaddress]{revtex4-1}
\usepackage{amsmath}    
\usepackage{graphicx}   
\usepackage{verbatim}   
\usepackage{color}      
\usepackage{subfigure}  
\usepackage{hyperref}   
\usepackage{multirow}
\usepackage{gensymb}
\newcommand{\siv}{SiV$^{0}$}
\newcommand{\sivm}{SiV$^-$}

\newcommand{\ti}{$T_1$}

\newcommand{\fig}[1]{{Fig.~#1}}

\makeatletter
\newcommand{\balancecolsandclearpage}{
  \close@column@grid
  \cleardoublepage
  \twocolumngrid
}

\begin{document}
\title{Optically detected magnetic resonance in neutral silicon vacancy centers in diamond via bound exciton states}
\author{Zi-Huai Zhang}
\affiliation{Department of Electrical Engineering, Princeton University, Princeton, New Jersey 08544, USA}

\author{Paul Stevenson}
\affiliation{Department of Electrical Engineering, Princeton University, Princeton, New Jersey 08544, USA}

\author{Gerg\H{o} Thiering}
\affiliation{Wigner Research Centre for Physics, P.O. Box 49, H-1525, Budapest, Hungary}
\affiliation{Department of Atomic Physics, Budapest University of Technology and Economics, Budafoki \'ut 8., H-1111 Budapest, Hungary}

\author{Brendon C. Rose} 
\affiliation{Department of Electrical Engineering, Princeton University, Princeton, New Jersey 08544, USA}

\author{Ding Huang} 
\affiliation{Department of Electrical Engineering, Princeton University, Princeton, New Jersey 08544, USA}

\author{Andrew M. Edmonds}
\affiliation{Element Six, Harwell, OX11 0QR, United Kingdom}

\author{Matthew L. Markham} 
\affiliation{Element Six, Harwell, OX11 0QR, United Kingdom}

\author{Stephen A. Lyon}
\affiliation{Department of Electrical Engineering, Princeton University, Princeton, New Jersey 08544, USA}

\author{Adam Gali}
\affiliation{Wigner Research Centre for Physics, P.O. Box 49, H-1525, Budapest, Hungary}
\affiliation{Department of Atomic Physics, Budapest University of Technology and Economics, Budafoki \'ut 8., H-1111 Budapest, Hungary}

\author{Nathalie P. de Leon}
\email{npdeleon@princeton.edu}
\affiliation{Department of Electrical Engineering, Princeton University, Princeton, New Jersey 08544, USA}

\date{\today}
\begin{abstract}
Neutral silicon vacancy (\siv{}) centers in diamond are promising candidates for quantum networks because of their excellent optical properties and long spin coherence times. However, spin-dependent fluorescence in such defects has been elusive due to poor understanding of the excited state fine structure and limited off-resonant spin polarization. Here we report the realization of optically detected magnetic resonance and coherent control of \siv{} centers at cryogenic temperatures, enabled by efficient optical spin polarization via previously unreported higher-lying excited states. We assign these states as bound exciton states using group theory and density functional theory. These bound exciton states enable new control schemes for \siv{} as well as other emerging defect systems.
\end{abstract}
\maketitle

Point defects in solid-state materials are promising candidates for quantum memories in a quantum network. These quantum defects combine the excellent optical and spin properties of isolated atoms with the scalability of solid-state systems \cite{Hanson2015Review,Wrachtrup2018Review,Awschalom2018Review}. Long-range, kilometer-scale entanglement generation has been demonstrated with the nitrogen vacancy (NV) center in diamond \cite{NVBelltest}. However, the entanglement generation rate in such demonstrations is limited by the optical properties of the NV center, which exhibits significant spectral diffusion \cite{Wolters2013,Chu2014} and a small Debye-Waller factor \cite{Barclay2011}. The neutral silicon vacancy center in diamond (\siv{}) has the potential to mitigate many of these problems; its inversion symmetry guarantees a vanishing permanent dipole moment, which minimizes spectral diffusion, and over 90$\%$ of its emission is in the zero-phonon line (ZPL) \cite{Rose2018}. However, there has been no report of optically detected magnetic resonance (ODMR) for this defect, a key first step towards establishing a spin-photon interface, and the electronic structure of \siv{} is still not well understood \cite{Green2019}. A detailed understanding of the optical transition and excited state structure of \siv{} is key in developing preparation, manipulation and readout schemes for quantum information processing applications.

In this work, we present the observation of previously unreported optical transitions in \siv{} that are capable of efficiently polarizing the ground state spin. Previous studies on \siv{} have reported a strong ZPL transition at 946\,nm, and a weaker strain-activated transition at 951\,nm \cite{Green2019}. Through a combination of optical and electron spin resonance (ESR) measurements, we are able to assign groups of transitions from 825 to 890~nm to higher-lying excited states of \siv{}. We interpret these spectroscopic lines as transitions to bound exciton (BE) states of the defect. We observe highly efficient bulk spin polarization through optical excitation of these transitions, providing another manifold of states that can be used for spin initialization. Spin polarization via these BE states while collecting emission from the ZPL and phonon sideband enables the observation of ODMR. We use ODMR measurements to probe the low magnetic field behavior of \siv{} where we observe no spin relaxation ($T_1$) out to 30~ms, spin dephasing times ($T_2^*$) of 202~ns, and spin coherence times ($T_2$) of 55.5~$\mu$s at 6~K. 

\begin{figure}[h!]
  \centering
  \includegraphics[width = 8.6cm]{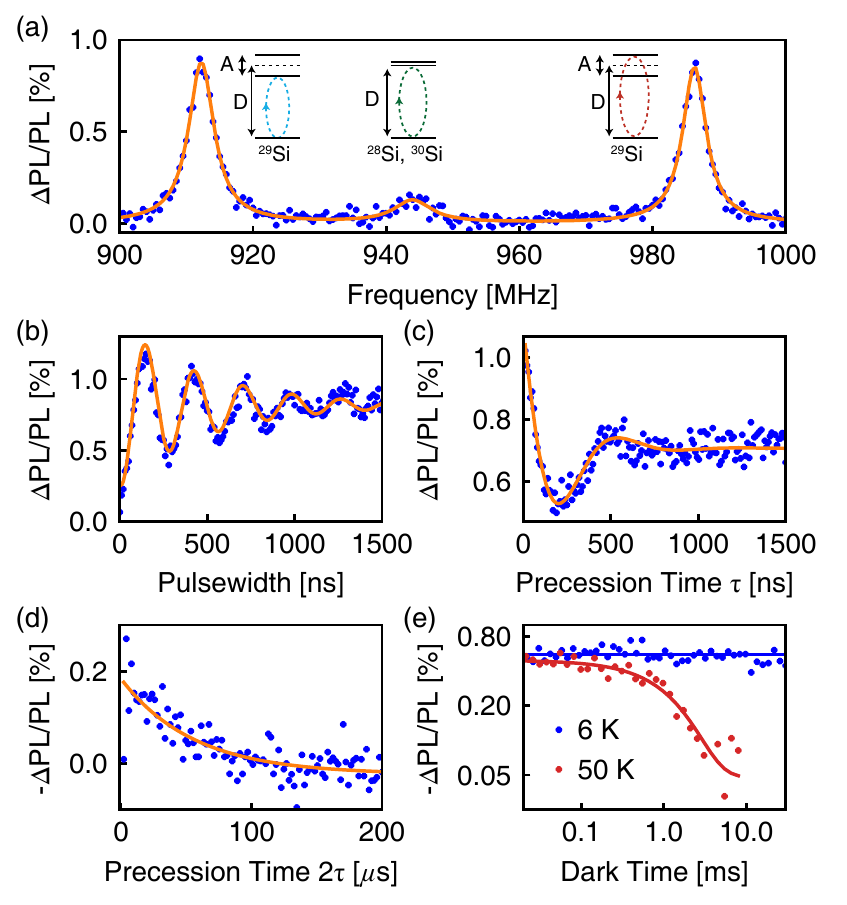}
  \caption{{\bf Optically detected magnetic resonance and coherent properties of \siv{} spins.}
  (a) CW ODMR spectrum measured at 6~K. The two outer peaks correspond to two hyperfine lines associated with the $^{29}$Si nucleus and the central peak is associated with $^{28}$Si and $^{30}$Si species. The solid line is a Lorentzian fit and the linewidths are microwave power broadened. Inset: relevant energy levels for individual spin transitions, where D denotes the zero-field splitting and A denotes the hyperfine interaction from the $^{29}$Si nucleus. (b) Rabi oscillation measured at 6~K performed at the lower hyperfine transition at 912~MHz. The data is fitted using $a\times e^{-t/T}\cos(\omega t + b)+c$ with $T =$ 499 $\pm$ 28~ns. (c) Spin dephasing time ($T_2^*$) measured at 6~K using a Ramsey sequence with microwave frequency detuned from the spin resonance by 1.6~MHz. The decay is fitted using $a\times e^{-\tau/T_2^*}\cos(\omega \tau + b)+c$ with $T_2^*$ = 202 $\pm$ 16~ns. (d) Spin coherence time ($T_2$) measured at 6~K with a Hahn echo sequence. The decay is fitted using $a\times e^{-2\tau/T_2}+b$ with $T_2$ = 55.5 $\pm$ 10.6~$\mu$s. The relatively large fitting error is due to the partially resolved modulation (see Supplemental Material Sec.~III\,A~\cite{Supplemental}). (e) Spin relaxation times ($T_1$) measured at 6~K and 50~K. At 6~K, no decay is observed up to 30~ms. The blue line is a flat line as a guide to the eye. At 50~K, we observe an exponential decay with a decay constant 1.38 $\pm$ 0.21~ms. The red line is a fit to the data with the form $a\times e^{-t/T_1} + b$. ODMR measurements are performed at ambient magnetic field.}
  \label{fig:Fig1}
\end{figure}

We observe ODMR in an ensemble of \siv{} centers using excitation at one of the BE transitions (855.65~nm) in a  chemical-vapor deposition grown sample doped with isotopically enriched $^{29}$Si during growth, described previously in Ref.~\cite{Rose60}. As the microwave frequency is swept across the zero-field splitting of \siv{}, we observe three resonance peaks in continuous-wave (CW) ODMR [\fig{\ref{fig:Fig1}(a)}]. The two outer peaks correspond to spin transitions associated with centers containing $^{29}$Si, while the central peak at 944 MHz is associated with $^{28}$Si and $^{30}$Si. The position and splitting of the lines are consistent with previously measured hyperfine parameters \cite{Edmonds2008}. 

We realize coherent control using pulsed ODMR on the lower frequency $^{29}$Si hyperfine transition at 912 MHz and observe Rabi oscillations that decay over 499 $\pm~28$~ns [\fig{\ref{fig:Fig1}(b)}]. We measure the spin dephasing time to be  $T_2^* = 202 \pm 16$~ns [\fig{\ref{fig:Fig1}(c)}] using a Ramsey sequence. By using a Hahn echo sequence to refocus the coherence, we measure the spin coherence time to be $T_2 = 55.5 \pm 10.6$~$\mu$s [\fig{\ref{fig:Fig1}(d)}]. The spin coherence time measured here is shorter than previous measurements of this sample using $X$-band pulsed ESR, $T_2 = 280-480$~$\mu$s \cite{Rose60}. This likely arises from the high density of \siv{} centers in this sample, which gives rise to instantaneous diffusion~\cite{Lyon2012,Rose60}. At ambient magnetic fields, the effects of instantaneous diffusion are more pronounced because centers of different orientations and nuclear spin projections 
are nearly degenerate. This effect limits  $T_2$ to 56~$\mu$s (see Supplemental Material Sec.~III\,B~\cite{Supplemental}).

We measure the spin relaxation time ($T_1$) using pulsed ODMR by measuring spin population decay after a variable dark time between the initialization and readout pulses. We observe no decay up to 30~ms at 6~K [\fig{\ref{fig:Fig1}(e)}], consistent with previous measurements of $T_1 = 46$~s at this temperature~\cite{Rose60}. At higher temperatures, the spin lifetime shortens significantly due to an Orbach process with an activation energy of 16.8~meV \cite{Rose60} and we measure $T_1 = 1.38 \pm 0.21$ ms at 50 K. 

Our temperature-dependent ODMR \ti{} measurements on the lower hyperfine transition are consistent with the previously measured activation energy (see Fig.~S7~\cite{Supplemental}), but we observe the Orbach rate prefactor to be $\sim$260 times larger. This is largely due to hyperfine-induced mixing of the \siv{} spin states (see Supplemental Material Sec.~III\,C~\cite{Supplemental}). The hyperfine interaction for \siv{} is $\sim$30 times larger than that for the NV center and the zero-field splitting is three times smaller~\cite{Edmonds2008,NVhyperfine2009}, so at low magnetic field the influence of the hyperfine interaction is much more pronounced. Unlike nitrogen, however, silicon has spin-free nuclear isotopes which may be used to circumvent these effects.

\begin{figure}[h!]
  \centering
  \includegraphics[width = 8.6cm]{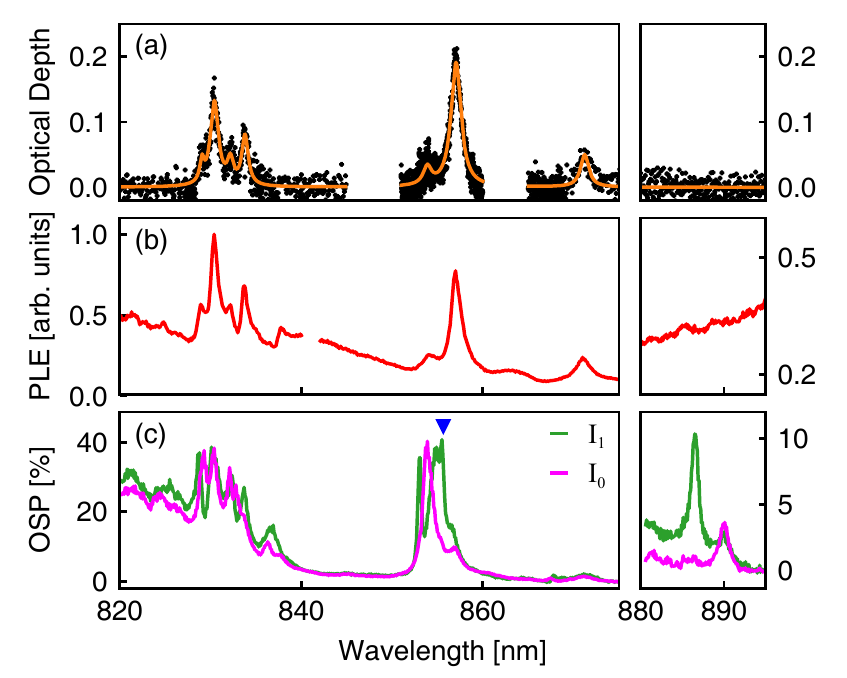}
  \caption{{\bf Spectroscopy and spin polarization of higher-lying excited states.}
  (a) Absorption measurements at 5.5~K showing narrow absorption peaks. The raw data is baseline subtracted to eliminate contribution from broadband absorption. The orange curve shows a Lorentzian fit to the data. (b) PLE measurement at 5.5~K with detection at 946~nm showing resonant features that line up with the observed absorption peaks. (c) Pump-probe OSP measurement at 5.5~K showing narrow resonances after initialization into $m_s = 0$ ($I_0$, magenta) or $m_s = 1$ ($I_1$, green). The amplitude of each spectrum $I_i$ represents probe induced population change of sublevel $m_s = i$, with the baseline subtracted. The blue triangle denotes the wavelength used for ODMR measurements. The wavelength range from 875 to 880~nm is not shown.} 
  \label{fig:Fig2}
\end{figure}

The observation of ODMR in \siv{} is enabled by the discovery of additional higher-lying excited states beyond the ZPL. Previous studies on \siv{} excited states were limited to the $^3E_u$ (ZPL at 946~nm) and $^3A_{2u}$ (ZPL at 951~nm) states but higher energy states were never explored. Transitions between 820 and 950~nm in silicon-doped diamonds have been previously observed with photoconductivity and absorption measurements, but there has been no detailed spectroscopy of these spectral lines, nor assignment of their microscopic origin \cite{Allers1995,Ulrica2010,Ulrica2011}.
    
In order to probe whether these transitions are associated with the \siv{} center, we correlate several types of optical spectroscopy at low temperature (5.5~K) at ambient magnetic field. First we perform absorption spectroscopy over a large wavelength range, from the ionization threshold ($\sim$826~nm \cite{Allers1995}) to 900~nm. We observe several families of peaks near 830, 855, and 870~nm [\fig{\ref{fig:Fig2}(a)}]. Then we perform photoluminescence excitation (PLE) spectroscopy, wherein we excite at these absorption wavelengths and detect emission at 946~nm, the ZPL of \siv{}. We observe the same clusters of resonances in PLE, confirming that the transitions are associated with the \siv{} center [\fig{\ref{fig:Fig2}(b)}]. 

Finally, we probe the interaction between these higher lying transitions and the ground state spin of \siv{} by measuring optical spin polarization (OSP) in bulk ESR ($\sim$3100~G) after excitation at these wavelengths [\fig{\ref{fig:Fig2}(c)}]. Specifically, we use a pump-probe measurement to isolate the contributions from $m_s = 0$ ($I_0$) and $m_s = 1$ ($I_1$) spin states (see Supplemental Material Sec.~VI~\cite{Supplemental}). Remarkably, the bulk OSP reaches values up to $40\% - 60\%$ (see Supplemental Material Sec.~V~\cite{Supplemental}), a key enabling capability for the observation of ODMR.

Using OSP measurements, we also observe a new cluster of transitions near 886 nm that are not evident in absorption or PLE spectroscopy [\fig{\ref{fig:Fig2}(c)}, right]. This indicates that these transitions have a weak oscillator strength, but are strongly spin polarizing.

\begin{figure}[htbp]
  \centering
  \includegraphics[width = 8.6cm]{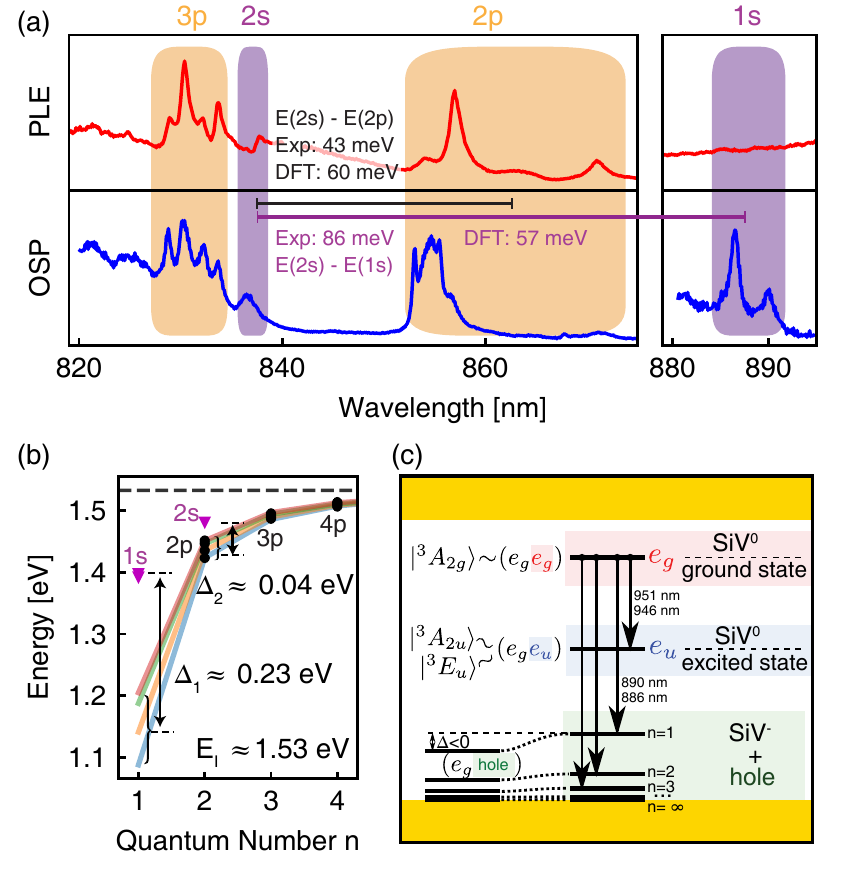}
  \caption{{\bf $s$- and $p$-like states in PLE and ESR measurements.}
  (a) State assignments and comparison of experimental and calculated energy splittings. The upper panel shows PLE spectra. The lower panel is constructed using decomposed OSP spectra as $2I_1+I_0$ to resemble absorption.  (b) Scaling of the peak positions extracted from PLE in (a). The fit uses Rydberg scaling $E_n = E_I - E_y/n^2$. Because of similar fine structures of $2p$ and $3p$ states, we fit different fine structure transitions separately corresponding to the different colored curves. The fitted ionization energy ($E_I$) and Rydberg energy ($E_y$) are 1.53 and 0.4 eV, respectively. The horizontal dashed line indicates the fitted ionization energy. States with $s$-like character are taken from spin-polarization measurements, and are shown with triangles. $\Delta_1$ and $\Delta_2$ are energy deviations for $1s$ and $2s$ states compared to the fitted Rydberg scaling that involve both central cell correction and the localized phonon energy. (c) Proposed bound exciton model for the higher-lying excited states showing orbital ground and excited states and BE states at higher energies in the hole picture. The 
 lower levels closer to the valence band maximum for electrons require higher excitation energy.} 
  \label{fig:Fig3_breakout}
\end{figure}

The number of observed transitions cannot be described by models utilizing only the orbitals localized on the \siv{} center. Group theoretic considerations describe three triplet excited configurations for \siv{}: $^3E_u$, $^3A_{1u}$ and $^3A_{2u}$~\cite{Gali2013}. Bulk photoluminescence measurements under uniaxial stress suggest that the 946~nm transition arises from the $^3E_u$ state and the 951~nm transition arises from the $^3A_{2u}$ state \cite{Green2019}. Only the transition from the $^3A_{1u}$ state has not been experimentally identified. 

The proximity of several of these resonances to the ionization threshold of \siv{} ($\sim$826~nm \cite{Allers1995}) provides a clue to their nature. We propose that \siv{} can act as a pseudo-acceptor, forming BE states composed of a hole weakly bound to a transiently generated \sivm{} center. BE states of neutral defects have been observed in SiC \cite{Egilsson1999,SiC2001}, Si \cite{Si1985Pline,Si1988Cline,Si1990,Si1994,Si1994Silver}, and GaP \cite{GaP1993}. One manifestation of BE states is a progression of peaks that can be described qualitatively as transitions between hydrogenic states and labeled with principal quantum numbers, $n$, and angular momentum labels ($s$, $p$, $d$, etc.). These progressions are observed in both PLE and OSP measurements, shown in \fig{\ref{fig:Fig3_breakout}(a)}. A schematic level diagram for the states described here is depicted in \fig{\ref{fig:Fig3_breakout}(c)}. Based on this model, transitions to ``$s$''-like states are expected to be electric-dipole forbidden, since both the ground state and BE state are of $gerade$ symmetry. Indeed, we observe transitions at 886 and 837~nm in OSP, but not in absorption or PLE. The isotopic shift of the $1s$ transition suggests that this transition is phonon assisted in nature (see Supplemental Material Sec.~VII~\cite{Supplemental}). We fit the observed energies ($E_n$) of the ``$p$''-like transitions to the Rydberg scaling, $E_n = E_I - E_y/n^2$, shown in \fig{\ref{fig:Fig3_breakout}(b)}, where $E_I$ is the ionization energy and $E_y$ is the Rydberg energy. We find the fitted ionization energy $E_I$ to be in good agreement with photoconductivity measurements \cite{Allers1995}, and the Rydberg energy to be consistent with an effective-mass description of the system (see Supplemental Material Sec.~VIII\,A~\cite{Supplemental}).

The $s$-like states were excluded from this analysis because of their vibronic nature and the central-cell correction expected for these types of states~\cite{cardona2005fundamentals}. 
This expectation is borne out in density functional theory (DFT) calculations (see Fig.~S20 and Supplemental Material Sec.~IX\,G~\cite{Supplemental}), where the calculated $1s$-$2s$ energy difference of 57~meV is in better agreement with experimental measurements (86~meV) than the $>250$~meV difference expected from a hydrogenic model without a central cell correction. The calculated energy difference between the $2s$ and $2p$ states is also consistent with experimental observations [\fig{\ref{fig:Fig3_breakout}(a)}]. The central cell correction arises because the BE states are effectively excluded from occupying the 6 carbon atoms adjacent to the \sivm{} center, increasing the effective Bohr radius and decreasing the effective Rydberg energy. This effect is less pronounced for $p$-like states because they have radial nodes at the \sivm{} center.

Within each labeled manifold in \fig{\ref{fig:Fig3_breakout}(a)}, significant structure is observed. This likely arises from a combination of spin-orbit structure in the valence band, crystal-field interactions from the presence of the symmetry-lowering \siv{} defect, and coupling between the bound hole and \sivm{}. We note that the bulk inhomogeneous linewidth likely obscures the full multiplicity of these transitions. 

Transitions above the $n=3$ level are not clearly observable in the experimental data. We believe this is a combination of the oscillator strength scaling ($n^{-3}$), proximity to the ionization threshold, and competition with other nonradiative, non-spin-polarizing relaxation pathways.

\begin{figure}[!htbp]
  \centering
  \includegraphics[width = 8.6cm]{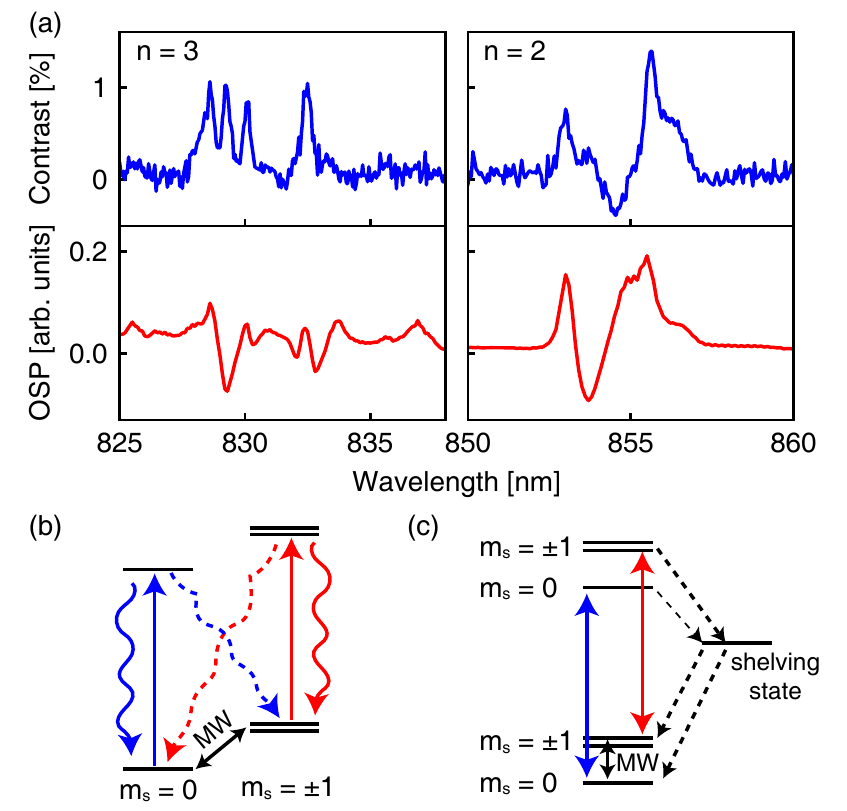}
  \caption{{\bf Wavelength dependence of ODMR and mechanisms for ODMR contrast.} (a) Upper panel: CW ODMR contrast as a function of excitation wavelength measured at 60~K. Microwave frequency is fixed at the lower hyperfine transition. Lower panel: OSP as a function of excitation wavelength at 5.5~K. ODMR contrast is measured at ambient magnetic field while OSP measurements are performed at $X$-band frequencies ($\sim$3100~G). (b) Level scheme for ODMR based on selective excitation of optical transitions with nonunity cyclicity. Dashed arrows indicate spin-non-conserving decay paths. (c) Level scheme for ODMR based on spin-dependent shelving. Dashed arrows represent nonradiative decay. Solid arrows indicate radiative transitions. MW denotes microwave driving.} 
  \label{fig:Fig4}
\end{figure}

With this model for the nature of the transitions, we now turn to the details of the spin polarization and ODMR contrast. The magnitude of the ODMR signal depends sensitively on the excitation wavelength, and we observe resonant features that match the linewidths observed in absorption, PLE, and OSP measurements for the $n=2$ and $n=3$ BE transitions [\fig{\ref{fig:Fig4}}(a), upper panel]. This is in stark contrast to ODMR in the NV center, which shows significant ODMR contrast for off-resonant excitation due to its spin dependent intersystem crossing. This indicates that the mechanism for ODMR relies on selective excitation of these transitions, which can arise from both the resonant nature of OSP and spin-selective optical pumping leading to population shelving into a ``dark" spin state. 

Furthermore, we observe that the ODMR signal can be both positive and negative. Optical transitions with nonunity cyclicity lead to population of ground states (e.g., other $m_s$ levels here) that are not addressed by the spectrally narrow excitation [\fig{\ref{fig:Fig4}}(b)]. This process has no preferential direction of spin-polarization (addressing different optical transitions may result in net polarization in either $m_s=0$ or $m_s=\pm 1$), but should result in positive contrast (brighter emission) under resonant microwave driving, as population is restored to the state being addressed by the optical excitation.

Another possible mechanism involves spin-dependent shelving of population in the excited state into a metastable state, which then decays back to the ground state [\fig{\ref{fig:Fig4}}(c)]. This mechanism is observed in the NV center under off-resonant excitation at room temperature. Here, the excitation addresses all spin sublevels in the ground state, and the different branching ratios in the excited state for different spin projections result in a spin polarization direction independent of excitation wavelength~\cite{Robledo_2011}. The sign of the ODMR contrast, however, has no such general restriction, and should depend on the specific details of the excited state manifold. 

We compare the OSP and the ODMR contrast for the $n=2$ and $n=3$ BE transitions in \fig{\ref{fig:Fig4}(a)}. Spin polarization both \textit{into} and \textit{out of} the $m_s=0$ state is observed, depending on the excitation wavelength. This suggests that optical pumping plays a role in the excitation cycle of these transitions. The ODMR contrast data, however, reveals that this is not a complete description. Although the $n=3$ data shows primarily positive contrast (brighter emission), the $n=2$ data shows clear negative contrast for some excitation wavelengths. This suggests that decay from the excited state into a different manifold of states is involved. 

In conclusion, we report the first realization of ODMR in \siv{} centers in diamond. We demonstrate coherent control of an ensemble of \siv{} spins at low magnetic field and measure $T_1$ much longer than 30~ms and $T_2$ of 55.5~$\mu$s at 6~K. ODMR is enabled by newly discovered higher-lying excited states of \siv{}, which allow for efficient optical spin polarization. We propose that these transitions arise from BE states, and we provide DFT calculations for the ionization threshold, central cell correction, and energy splitting between different states that are consistent with experimental observations. On-going work includes single center ODMR measurements, as well as investigating the microscopic mechanism for ODMR via BE states. Our measurements indicate that ODMR cannot arise solely from spin-dependent shelving of population or resonant optical pumping into a dark state, and it is likely that a combination of processes give rise to the observed features. 

Optical spin polarization via these BE states enables a powerful method of spin initialization and readout for \siv{} centers in diamond. In particular, their resonant nature allows for the use of much lower excitation powers, which circumvents optically induced noise from the bath \cite{Siyushev2013}. More broadly, this scheme can potentially be deployed in other emerging defect systems, such as other neutral group IV vacancy centers in diamond \cite{Thiering2018PRX,Thiering2019} and neutral divacancy centers in SiC \cite{RN51}.

\vskip 0.2in
We thank J.~Thompson for fruitful discussions, as well as S.~Kolkowitz, L.~Rodgers, and Z.~Yuan for comments on the manuscript. This work was supported by the NSF under the EFRI ACQUIRE program (Grant No.~1640959) and through the Princeton Center for Complex Materials, a Materials Research Science and Engineering Center (Grant No.~DMR-1420541). This material is also based on work supported by the Air Force Office of Scientific Research under Grant No.~FA9550-17-0158, and was partly supported by DARPA under Grant No.~D18AP00047. G.~T. was supported by the J\'anos Bolyai Research Scholarship of the Hungarian Academy of Sciences and the \'UNKP-20-5 New National Excellence Program of the Ministry of Innovation and Technology in Hungary (ITM) from the National Research, Development and Innovation Office in Hungary (NKFIH). D.~H. was supported by a National Science Scholarship from A*STAR, Singapore. A.~G.\ acknowledges the support from NKFIH for Quantum Technology Program (Grant No.\ 2017-1.2.1-NKP-2017-00001) and National Excellence Program (Grant No.\ KKP129866), from the ITM and NKFIH for the Quantum Information National Laboratory in Hungary, from the EU Commission (Asteriqs project, Grant No.\ 820394) and the EU QuantERA program (Q\_magine project, NKFIH Grant No.\ 127889).

\bibliography{bibliography}
\nocite{*}

\balancecolsandclearpage
\widetext
\begin{center}
\textbf{\large Supplemental Material for \\``Optically detected magnetic resonance in neutral silicon vacancy centers in diamond via bound exciton states''}
\label{Sec:SI}
\end{center}

\setcounter{figure}{0}
\setcounter{section}{0}

\renewcommand{\thefigure}{S\arabic{figure}}
\renewcommand{\thetable}{\Roman{table}}
\renewcommand{\thesection}{\Roman{section}}
\renewcommand{\theHfigure}{Supplement.\thefigure}

\section{\label{SI_Setup} SUPPLEMENTARY EXPERIMENTAL METHODS}

\subsection{Experimental Setups}
\textbf{Sample preparation:}\quad Three different \{110\} diamonds grown by chemical vapor deposition were studied. The first two samples (D1 and D2) were doped during growth with silicon. The silicon precursor was isotopically enriched with 90\% ${^{29}}$Si (resulting in similar residual concentration of ${^{28}}$Si and ${^{30}}$Si). After high-pressure-high-temperature annealing, the \siv{} concentration is $4.0 \times 10^{16}$ cm$^{-3}$ for sample D1 \cite{Rose60}. Sample D2 was cut along the growth direction so its \siv{} concentration depends on the specific region under study. We estimate its \siv{} concentration to be $2.4 \times 10^{15}$ cm$^{-3}$ for the region studied in photoluminescence excitation (PLE) measurements. The third diamond (D3) was doped during growth with boron and implanted with ${^{28}}$Si, as described in \cite{Rose2018}. After annealing, the resulting \siv{} concentration in the implanted layer is $5.1 \times 10^{15}$ cm$^{-3}$. Sample D1 is studied in the main text while samples D2 and D3 are measured to provide additional data in the supplemental material. Sample D1 shows a preferential alignment of \siv{} such that the in-plane and out-of-plane sites have a density ratio of 1:3 \cite{Ulrica2011}.
\\

\textbf{Electron spin resonance (ESR):}\quad Pulsed X-band ($\sim$9.5~GHz) ESR is performed on a modified Bruker Elexsys 580 system using a dielectric volume resonator (Bruker MD5) and a 1.4-T electromagnet, the details of which are thoroughly described elsewhere \cite{Rose60}. Optical illumination is applied through a multi-mode fiber (Thorlabs FT400EMT) positioned above the sample. A narrow linewidth tunable CW Ti:Sapphire laser (Msquared SolsTis) is used as the excitation source for 800 nm - 1000 nm. For pump-probe ESR measurements, a second narrow linewidth tunable laser (Toptica CTL 950) is used as the pumping laser. All measurements are performed on the $m_s = 0 \leftrightarrow +1$ transition with the magnetic field aligned to a $\langle 111 \rangle$ direction of the sample unless otherwise noted. Optical spin polarization (OSP) is measured using a two-pulse Hahn echo sequence (200 ns $\pi$ pulse) after optical excitation. The echo intensity is normalized to the echo intensity resulting from thermal polarization in the dark. The sign of OSP is defined as the relative population of the spin levels, with positive (negative) OSP being more polarization into $m_s = 0$ ($m_s = 1$) state. OSP is measured on the $^{29}$Si hyperfine line for samples D1 and D2 unless otherwise noted. All ESR measurements are performed at 5.5~K.
\\

\textbf{Photoluminescence excitation (PLE):}\quad All optical measurements are performed in a helium flow cyrostat (Janis ST-100) with the sample mounted on a copper cold finger. Excitation and detection channels for PLE are separated by a dichroic beam splitter (Semrock FF924-Di01). Excitation is focused on the sample with a 30 mm doublet lens. Emission is further filtered with a tunable 937 nm long-pass filter (Semrock FF01-937/LP-25) and coupled to a 50 $\mu$m multimode fiber that routes the signal to a grating spectrometer (Princeton Instruments Acton SP-2300i). At each excitation wavelength, we acquire a photoluminescence (PL) spectrum and plot integrated emission at the 946 nm peak.
\\

\textbf{Absorption:}\quad For absorption measurements, the laser is split into two paths. One path travels through both the diamond and the windows of the cryostat while the other travels through only the windows of the cryostat, serving as a reference. Transmitted power through each path is measured with a Si photodiode (Thorlabs DET100A). The thickness of the diamond sample (D1) used for absorption is 0.66~mm.
\\

\textbf{Optically detected magnetic resonance (ODMR):}\quad For ODMR, the laser is coupled to an acousto-optic modulator (AOM, Isomet 1305C-1) for pulsed excitation. ODMR experiments use the same optical setup as PLE except that the signal is sent to a single photon detector (Excelitas SPCM-AQRH) and the excitation is focused on the sample with a 10X near infrared objective (Olympus LMPLN10XIR) outside of the cryostat. Microwave (MW) excitation is applied using a 70~$\mu$m wire stretched across the sample. The MW excitation is generated with a signal generator (Rohde and Schwarz SMATE200A) and then amplified by a high-power MW amplifier (Ophir 5144). Two 0.8 - 2~GHz MW circulators (Ditom D3C0802S) are added after the amplifier for circuit protection. The MW excitation is pulsed using a fast MW switch (Mini-Circuits ZASWA-2-50DR+). The timing of MW pulses and optical pulses are synchronized using a TTL pulse generator (SpinCore PBESR-PRO-500). A home-built Helmholtz coil is used to apply a magnetic field along one of the in-plane $\langle111\rangle$ directions. For time-tagging the photon counts, a time-correlated single photon counting system (PicoQuant PicoHarp 300) is used.

\subsection{ESR Pulse Sequences}
\begin{figure}[h]
  \centering
  \includegraphics[width = 0.5\textwidth]{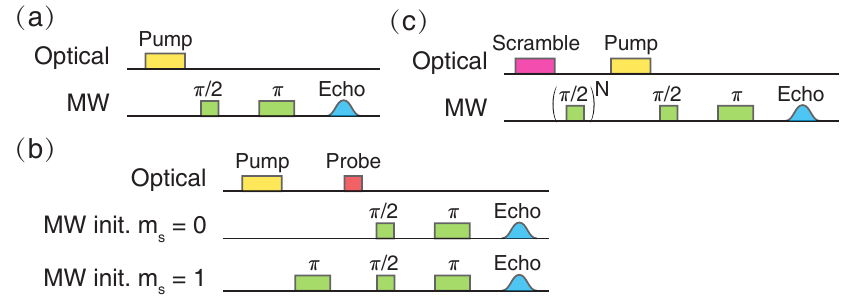}
  \caption{{\bf Pulse sequences for ESR.} (a) Echo detected optical spin polarization. (b) Pump-probe measurement with different spin initialization sequences. (c) Pulse sequence for measuring polarization saturation curves.} 
  \label{FigS3}
\end{figure}
In pulsed ESR, the echo intensity is proportional to the population difference of the two spin levels under study. Spin polarization is measured by monitoring the integrated echo intensity from a standard Hahn-echo sequence after an optical pump pulse [\fig{\ref{FigS3}(a)}]. For state-resolved measurements, the spins are first initialized into $m_s = 0$ with a long optical pump pulse at 946.76 nm to achieve efficient OSP from ZPL excitation. For $m_s = 1$ initialization, a MW $\pi$ pulse is then applied to invert the population. After initialization, a short optical probe pulse is applied [\fig{\ref{FigS3}(b)}]. For data shown in 
Fig.~2(c), the length of the optical pump pulse (80~mW excitation power) is 6~s for measurements between 820~nm to 875~nm, and 4~s for measurements between 880~nm and 895~nm. The length of the optical probe pulse ($\sim$45~mW excitation power) is 100~ms for measurements between 820 nm and 875~nm, and 500~ms for measurements between 880~nm and 895~nm. Polarization saturation curves are measured by shining an optical pump pulse with different pulse lengths. To avoid waiting for the spins to reset after each pulse sequence, an off-resonant optical pulse and N evenly spaced $\pi/2$ pulses are applied to scramble the spin polarization before the pump pulse [\fig{\ref{FigS3}(c)}]. For these measurements, N=6. The analysis for these data is described in Section~\ref{SI:SaturationCurves} and Section~\ref{Sec:Spec_Decomp}.

\subsection{\label{ODMRseq}ODMR Pulse Sequences}

\begin{figure}[h]
  \centering
  \includegraphics[width = 0.5\textwidth]{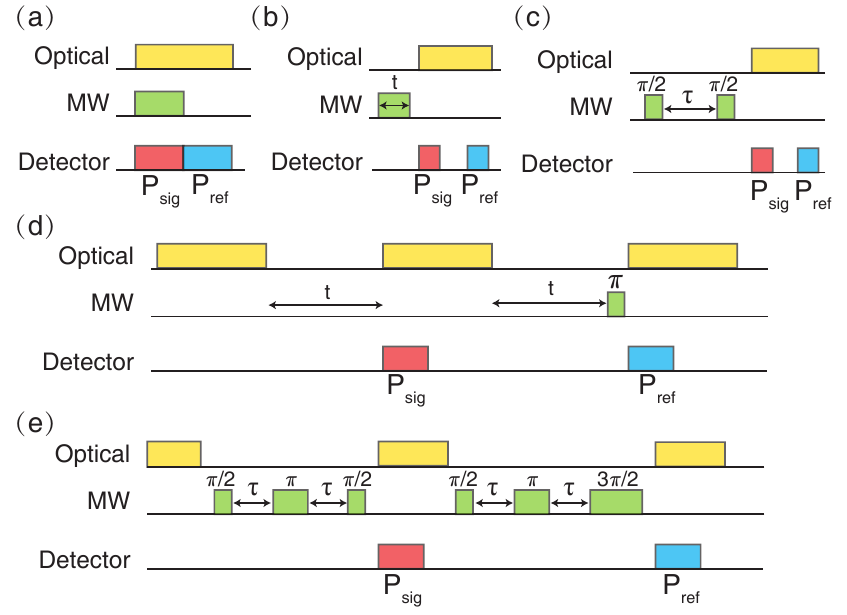}
  \caption{{\bf Pulse sequences for ODMR.} (a) Pulse sequence for CW ODMR. The switching rate for MW pulses is set to 1~kHz. (b) Pulse sequence for Rabi measurement. The MW pulses on resonance with spin transitions are applied with varying duration. (c) Ramsey sequence for $T_2^*$ measurement. The free precession time between two $\pi/2$ pulses is swept. (d) Pulse sequence for $T_1$ measurement. The phase of detection is alternated between $m_s = 0$ (no MW pulse) or $m_s = \pm1$ (with MW $\pi$ pulse). (e) Pulse sequence for $T_2$ measurement. The phase of detection is alternated between $m_s = 0$ (with MW $\pi/2$ pulse) or $m_s = \pm1$ (with MW $3\pi/2$ pulse).} 
  \label{FigS1}
\end{figure}

To suppress slow noise in continuous-wave (CW) ODMR measurements, we modulate the MW pulses on and off at a rate of 1~kHz. Photon counts are gated when the MW tone is on ($P_{sig}$) and off ($P_{ref}$), and ODMR contrast is normalized using $\Delta PL/PL = P_{sig}/P_{ref}-1$ [\fig{\ref{FigS1}(a)}]. 

For pulsed ODMR measurements, the large dynamic range of laser pulse duty cycle gives rise to systematic fluctuations in the laser power because of AOM heating. To correct for this effect, we use two types of normalization for pulsed ODMR experiments. For Rabi and $T_2^*$ measurements where the laser is gated mostly on, we use a standard detection scheme [\fig{\ref{FigS1}(b)} and  \fig{\ref{FigS1}(c)}]. Two 10 $\mu$s detection windows separated by 50 $\mu$s are applied during the readout pulse. The first window measures the transient spin population ($P_{sig}$) after the MW pulses while the second window measures the steady state spin population ($P_{ref}$). The normalized signal is calculated as $\Delta PL/PL = P_{sig}/P_{ref}-1$. 

For $T_1$ and $T_2$ measurements where the duty cycle varies significantly with delay time, we alternately apply different microwave pulses before the readout pulse to invert the phase of detection. For $T_1$, we alternate between applying a $\pi$ pulse and not applying any MW pulse in order to provide a reference count rate [\fig{\ref{FigS1}(d)}]. This also ensures the timescale we measure is related to the spin-relaxation, and does not include contributions from other optical processes. For $T_2$, we alternate between applying a $\pi/2$ pulse or a $3\pi/2$ pulse [\fig{\ref{FigS1}(e)}]. The data taken with phase inversion is normalized as $\Delta PL/PL = (P_{sig}-P_{ref})/(P_{sig}+P_{ref})$.

\subsection{Transient Spin-dependent Fluorescence}

\begin{figure}[h]
  \centering
  \includegraphics[width = 0.5\textwidth]{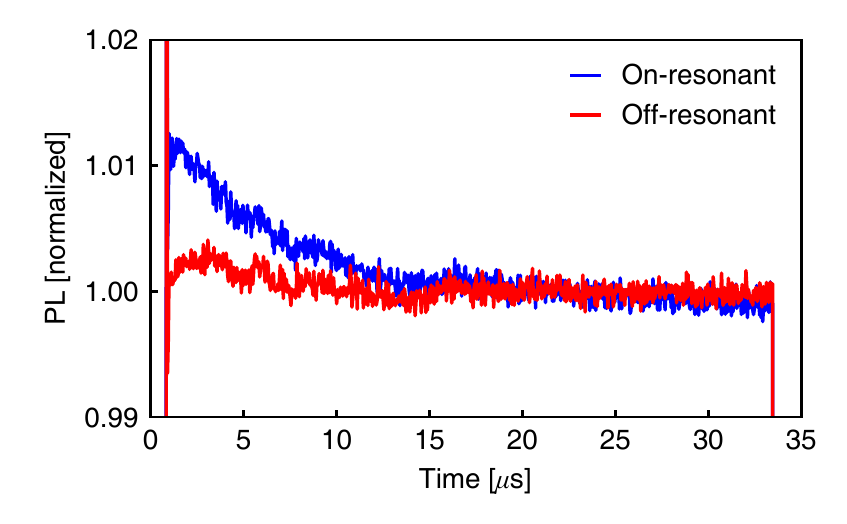}
  \caption{{\bf Transient PL response.} PL traces for different spin initialization. The spike at the beginning is from a pile-up artifact of the time-tagger due to the high photon rate compared to pulse repetition rate.} 
  \label{FigS2}
\end{figure}

To determine the optimum integration window for ODMR measurements, we measure transient spin-dependent PL by time-tagging the photon counts. A long optical pulse (30~$\mu$s) first polarizes the spin ensemble. Then we apply an on-resonant (off-resonant) MW $\pi$ pulse to flip (not flip) the spin state. The time traces for different spin states are shown in \fig{\ref{FigS2}}. We observe spin-dependent PL up to 15~$\mu$s. The integration windows and optical pulse duration are set accordingly: we set the spin polarization time to 75~$\mu$s, and we set the detection window to 10~$\mu$s. 

\section{Additional Characterization on Multiple Samples}

To confirm that the higher-lying excited states are a feature intrinsic to \siv{} centers rather than some sample dependent phenomenon, we measure OSP and PLE on two bulk-doped samples (D1 and D2) and a third implanted sample, D3.

\begin{figure}[!htbp]
  \centering
  \includegraphics[width = 0.5\textwidth]{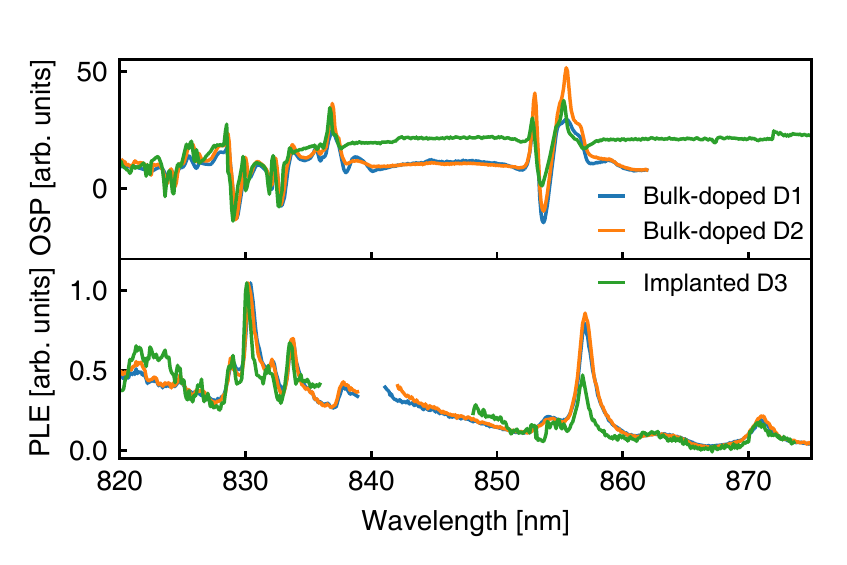}
  \caption{{\bf Higher-lying states in multiple samples.} Upper panel: OSP measured from the three samples. Lower panel: PLE measured from the three samples} 
  \label{FigS4}
\end{figure}

The spectra show consistent optical transitions and spin polarization behavior (\fig{\ref{FigS4}}). An isotopic shift is observed between D1, D2 ($^{29}$Si enriched) and D3 ($^{28}$Si implanted), arising from differences in the zero-point energy of the local phonons.

\section{\label{Sec:FieldDependence}Low-Field Spin Dynamics}
\subsection{Magnetic Field Dependence of ODMR Spectrum and Envelope Modulation}

\begin{figure}[t]
  \centering
  \includegraphics[width = 0.5\textwidth]{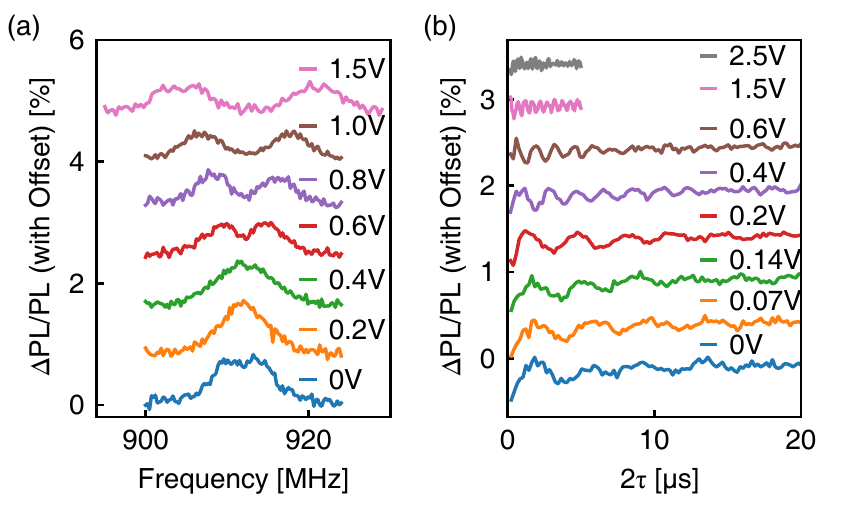}
  \caption{{\bf Magnetic field dependence of ODMR spectrum and $T_2$ modulation.} (a) ODMR spectrum for the lower hyperfine transition is measured at different magnetic fields. The peaks observed here come from the three off-axis sites ($\sim109.5\degree{}$ misalignment). The traces are intentionally offset with each other. (b) Echo decay envelope measured with fine time steps at different magnetic fields. Faster modulation is observed at higher applied magnetic field. 2$\tau$ is the total free precession time. The lower signal amplitude at higher magnetic fields is due to decreased microwave pulse fidelity because of more spatially inhomogeneous magnetic fields. The legends indicate the voltages applied for the Helmholtz coil. We find that a scaling factor of 6~G/V best reproduces our observations.} 
  \label{FigS12}
\end{figure}

A Helmholtz coil positioned along the in-plane $\langle 111 \rangle$ direction of sample D1 applies a small magnetic field. With this configuration the magnetic field is misaligned by $\sim109.5\degree{}$ with respect to three sites, which are therefore degenerate. Due to the lower concentration of defects oriented in-plane (Section \ref{SI_Setup}), significantly lower signal-to-noise ratio is expected for that site. As a result, we focus on the sites oriented $\sim109.5\degree{}$ to the field. Upon applying magnetic field, Zeeman splitting is observed, confirming the spin-dependent nature of these ODMR transitions [\fig{\ref{FigS12}}(a)]. The broadening of the lines at higher magnetic fields likely arises from a combination of inhomogeneity of the magnetic field for different sites and splitting of hyperfine transitions in the presence of an off-axis magnetic field.

\begin{figure}[]
  \centering
  \includegraphics[width = 0.5\textwidth]{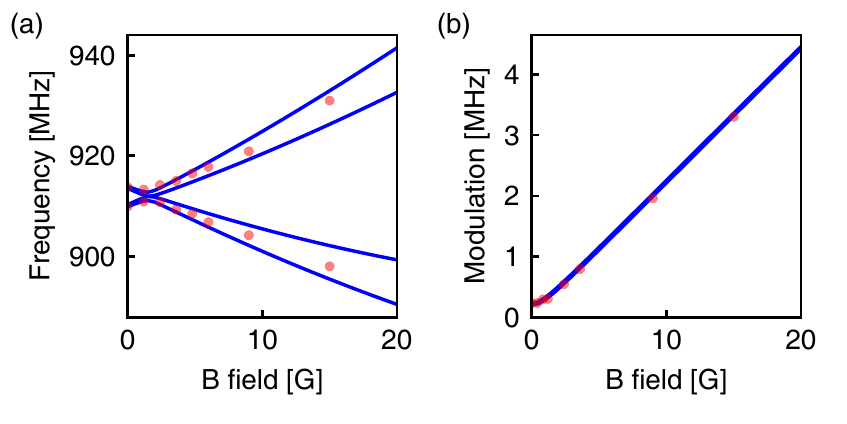}
  \caption{{\bf Magnetic field dependence of level splitting and modulation frequency.} (a) Simulated ODMR spectrum as a function of off-axis (113.76\degree{} misalignment) magnetic field (blue traces). We include a stray field term B = [0, 0.9, 0.65] G before applying any magnetic field. The red dots are experimental ODMR peak positions. Note that the closest pairs of simulated transitions are comparable the experimental linewidth and so cannot be resolved. (b) Modulation frequency as a function of magnetic field. The blue curve is calculated using the simulated level splittings. The red dots are experimental values. The applied magnetic field strength is estimated using the ODMR peak positions. We find that a scaling factor of 6~G/V best reproduces our data. We note that due to experimental limitations, we are only able to get a rough estimation of field alignment and strength based on ODMR peak positions.} 
  \label{FigS11}
\end{figure}

A pronounced modulation of the spin echo decay is observed in our data [\fig{\ref{FigS12}(b)}]. The observed oscillation frequency increases with magnetic field, and arises from a set of hyperfine transitions being driven simultaneously in our experiment. To probe this further, we simulated the expected ODMR spectrum at low magnetic fields. Four transitions are present in total [\fig{\ref{FigS11}(a)}], but the separations are often smaller than the linewidths measured from CW ODMR. The four transitions at zero applied field can be labeled approximately as (from lowest to highest frequency)
\begin{equation}\lvert -1\rangle \lvert \uparrow \rangle \leftrightarrow \lvert 0\rangle(\lvert \uparrow \rangle -i \lvert \downarrow \rangle), \end{equation}
\begin{equation}\lvert -1\rangle \lvert \uparrow \rangle \leftrightarrow \lvert 0\rangle(\lvert \uparrow \rangle +i \lvert \downarrow \rangle),\end{equation}
\begin{equation}\lvert 1\rangle \lvert \downarrow \rangle \leftrightarrow \lvert 0\rangle(\lvert \uparrow \rangle -i \lvert \downarrow \rangle),\end{equation}
\begin{equation}\lvert 1\rangle \lvert \downarrow \rangle \leftrightarrow \lvert 0\rangle(\lvert \uparrow \rangle +i \lvert \downarrow \rangle),\end{equation}
where the triplet electronic spin levels are labeled by $\lvert 1 \rangle$,$\lvert 0 \rangle$, $\lvert -1 \rangle$ and the nuclear spin levels are labeled by $\lvert \uparrow\rangle$ and $\lvert \downarrow\rangle$. During the free precession time of spin echo sequence, extra phase accumulates between two nearby hyperfine levels owing to their energy difference. We simulate the effect of this extra phase accumulation on spin echo using the rotating frame Hamiltonian $H = E_\Delta \lvert g_2\rangle \langle g_2 \lvert + \Omega_1 (\lvert g_1\rangle \langle e \lvert +\lvert e\rangle \langle g_1 \lvert) + \Omega_2 (\lvert g_2\rangle \langle e \lvert +\lvert e\rangle \langle g_2 \lvert)$ and find that the energy difference $E_\Delta$ between the two levels and the modulation frequency $f$ are related by $f = 0.5E_\Delta$. By measuring the magnetic field dependence of the modulation frequency, we find consistent results between experiment and simulation shown in \fig{\ref{FigS11}(b)}.

\subsection{\label{Sec:T2}Spin Coherence Times ($T_2$) on Sample D1}
The spin coherence time $T_2$ of \siv{} was previously characterized to be $\sim$1 ms below 20~K at X-band in sample D3, and was shown to be limited by spectral diffusion arising from the naturally abundant $^{13}$C bath. $T_2$ for sample D1 at X-band was extensively studied in Ref.~\cite{Rose60}. It was shown that $T_2$ for sample D1 is instead limited by instantaneous diffusion due to the high \siv{} concentration to be:
\begin{equation}
    \frac{1}{T_2} = \frac{1}{T_{2(SD)}} + \frac{1}{T_{2(ID)}},
\end{equation}
where $T_{2(SD)} = 0.95$~ms is the spectral diffusion-limited $T_2$ and $T_{2(ID)}$ is the instantaneous diffusion-limited $T_2$. The four orientations of \siv{} in D1 show preferential alignment with a population ratio of 1:1:3:3 \cite{Ulrica2011}. The two out of plane sites have 3 times higher \siv{} concentration compared to the two in-plane sites with $T_{2(ID)} = 0.319$~ms. For the higher (lower) concentration sites, $T_2$ was limited to 0.28~ms (0.48~ms). 

In the low magnetic field regime where we performed ODMR measurements, we could not isolate a single site or a single spin transition. Since instantaneous diffusion is proportional to the spin density, a shorter $T_2$ (limited by greater instantaneous diffusion) is expected. Driving all four sites leads to a factor of 8/3 increase in \siv{} concentration. Another factor of 2 is expected from the fact that X-band measurements address a single nuclear spin level while at low field, we address both nuclear spin levels simultaneously. These two factors together lead to an instantaneous diffusion limited $T_{2(ID)}$ of $0.319 \times 3/8 \times 1/2 = 0.06$ ms and $T_2 \approx 56$ $\mu$s. We note that differences in optical spin polarization and MW pulse fidelity between ODMR and X-band ESR are not considered in the estimation here, which could also affect the total spin density under MW driving.

\subsection{\label{T1orbach}Temperature Dependence of Spin Relaxation Times at Low Magnetic field}
\begin{figure}[h!]
  \centering
  \includegraphics[width = 0.5\textwidth]{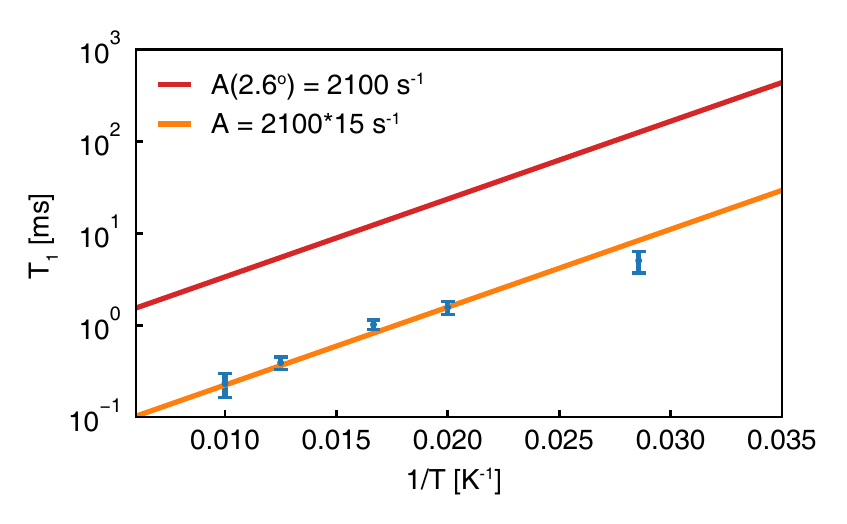}
  \caption{{\bf Temperature dependence of $T_1$ near zero field.} Red line represents the fitted Orbach process limited $T_1$ measured at X-band when the magnetic field is $2.6\degree$ misaligned to the defect axis. Orange line is for guidance with a 15 times larger pre-factor for the Orbach process.} 
  \label{FigS9}
\end{figure}

Spin relaxation times ($T_1$) are measured at different temperatures for the lower hyperfine transition at low magnetic field for sample D1. We note that temperature dependence of $T_1$ for \siv{} was previously studied using X-band ESR and can be described with an Orbach process~\cite{Rose60}
\begin{equation}
\frac{1}{T_{1}} = \frac{1}{T_{1sat}} + A(\theta)e^{-E_a/k_bT},
\end{equation}
where $T_{1sat}$ is the saturated $T_1$ at low temperature, $E_a$ is the activation energy for the Orbach process and $A(\theta)$ is a prefactor that depends on the misalignment $\theta$ between the defect axis and the spin axis.

We find that the measured $T_1$ here is about 15 times shorter compared to the X-band measurement for sample D1 ($\theta = 2.6\degree$) and 105 times shorter compared to the X-band measurement for sample D3 ($\theta = 0.8\degree$) but still follows an exponential scaling with increasing temperature. The similar exponential dependence suggests that $T_1$ is likely also limited by Orbach process. The prefactor $A(\theta)$ was shown to be strongly anisotropic due to different mixing rates between different spin states and phonon-activated Orbach excited state. Normally, when the magnetic field is aligned with the defect axis, no mixing of spin levels should occur so there shouldn't be any magnetic field dependence of the anisotropy. However, for the $^{29}$Si enriched sample, we must also consider mixing of the Zeeman states due to the hyperfine interaction. At X-band, the transverse hyperfine interaction ($\sim$79~MHz) is small compared to Zeeman splitting ($\sim$9.5~GHz) so it can be ignored. At low magnetic field, the transverse hyperfine interaction is significant compared to the zero-field splitting (942~MHz) and non-negligable mixing occurs (Table~\ref{TabS1}). For the lower hyperfine transition being measured here, we estimate using a Wigner rotation matrix $R(\theta) = e^{i\theta S_y}$ that at zero field, the hyperfine interaction induced mixing is equivalent to a $\theta \sim$5.0\degree{} rotation of the spin basis. 

\begin{table}[h!]
\caption{\label{TabS1}{Eigenstates for $^{29}$\siv{} at zero magnetic field. Each row represents the eigenstate for a specific eigenenergy. The triplet electronic spin levels are labeled by $\lvert 1 \rangle$, $\lvert 0 \rangle$ and $\lvert -1 \rangle$ while the nuclear spin levels are labeled by $\lvert \uparrow\rangle$ and $\lvert \downarrow\rangle$. Without hyperfine interaction, $\lvert \pm1 \rangle$ states are 0.94~GHz higher in energy compared to the $\lvert 0 \rangle$ state. The negative energy for the lowest eigenstates here is due to hyperfine induced mixing.}}
\begin{tabular}{lllllll}
\hline \hline
Energy (MHz) & $\lvert 1\uparrow\rangle$ & $\lvert 1\downarrow\rangle$ & $\lvert 0\uparrow\rangle$ & $\lvert 0\downarrow\rangle$ & $\lvert -1\uparrow\rangle$ & $\lvert -1\downarrow\rangle$
\tabularnewline
\hline 
980.15 & 1 & 0 & 0 & 0 & 0 & 0
\tabularnewline
980.15 & 0 & 0 & 0 & 0 & 0 & 1
\tabularnewline
907.28 & 0 & 0.998 & 0.061 & 0 & 0 & 0
\tabularnewline
907.28 & 0 & 0 & 0 & 0.061 & 0.998 & 0
\tabularnewline
-3.43 & 0 & 0 & 0 & 0.998 & -0.061 & 0
\tabularnewline
-3.43 & 0 & -0.061 & 0.998 & 0 & 0 & 0
\tabularnewline
\hline \hline
\end{tabular}
\end{table}

We estimate the reduction in \ti{} by calculating the ratio between $A(0\degree)$ and $A(\theta)$ at the experimental misalignments using the parameters determined in Ref.~\cite{Rose60}. For the previous X-band measurements, $A(0.8\degree)/A(0\degree) = 2.52$ and  $A(2.6\degree)/A(0\degree) = 17.04$. This is consistent with the experimentally determined $A(2.6\degree) = 2.10\times10^3 ~s^{-1}$ for sample D1 and
$A(0.8\degree) = 3\times10^2 ~s^{-1}$ for sample D3. For the $5\degree$ misalignment caused by hyperfine interaction at zero-field, $A(5.0\degree)/A(0\degree) = 60$. This accounts for most of the observed reduction ($A_{exp}/A(0\degree) \sim 17.04 \times 15 \sim 256$) in \ti{} compared to a perfect $0\degree$ misalignment.

By inspecting the eigenstates from Table~\ref{TabS1}, the 980.15 MHz levels (which are involved in the higher hyperfine transition) are not mixed by the transverse hyperfine interaction: they remain pure $\lvert \pm 1\rangle$ states. The $T_1$ anisotropy of \siv{} was modeled by extracting a larger $m_s=0$ state overlap with the Orbach excited state \cite{Rose60}. Therefore, we expect the higher hyperfine transition to have longer \ti{}.

\subsection{\label{ContrastTdep}Temperature Dependence of ODMR Contrast}
\begin{figure}[h]
  \centering
  \includegraphics[width = 0.7\textwidth]{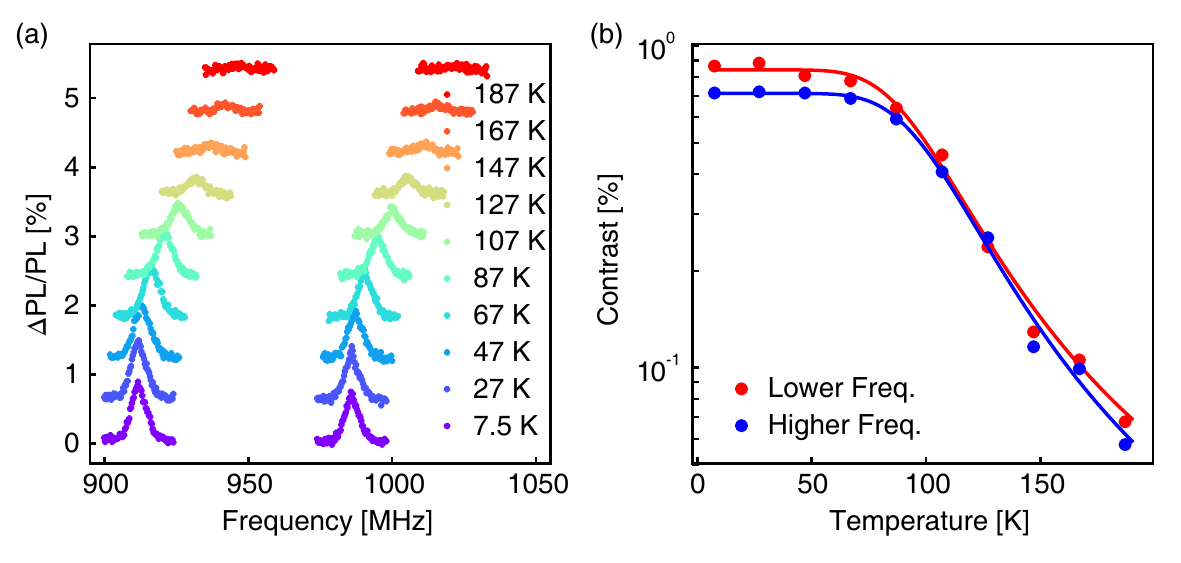}
  \caption{{\bf Temperature dependence of ODMR contrast.} (a) Continuous-wave ODMR spectra measured at different temperatures using $\sim$855.65~nm excitation. The plots are offset with each other for comparison. (b) Temperature dependence of ODMR contrast extracted by performing Lorentzian fits to the peaks in (a). The solid lines are fits using $1/C(T) = 1/C_{sat} + A\times e^{-E_b/k_BT}$.} 
  \label{FigS17}
\end{figure}

We measure the temperature dependence of continuous-wave ODMR spectra on sample D1 [\fig{\ref{FigS17}(a)}]. ODMR contrast is flat up to 70~K, and is still observable (about 0.07\%) at 187~K. While the underlying physical process requires further detailed study, we fit the temperature dependence of contrast using a phenomenological model
\begin{equation}
\frac{1}{C(T)} = \frac{1}{C_{sat}} + A\times e^{-E_b/k_BT},
\end{equation}
where $C_{sat}$ is the saturated contrast at low temperatures, A is an amplitude prefactor, T is the temperature, $k_B$ is the Boltzmann constant, and $E_b$ is the activation energy of a phonon-activated process. The fits yield an activation energy of $E_b = 50.8 \pm 4.6$~meV for the lower hyperfine line, and $E_b = 56.3 \pm 2.6$~meV for the higher hyperfine line. This temperature dependence differs from that of $T_1$ (16.8 meV activation energy), which suggests that shorter $T_1$ at higher temperature alone cannot explain the observed temperature dependence. Therefore, we suspect that additional phonon-induced spin mixing, and phonon-mediated non-radiative decay from the bound exciton excited states are responsible for the observed temperature dependence. 

\subsection{\label{ODMRPdep}Power Dependence of ODMR Spectra}
\begin{figure}[htbp!]
  \centering
  \includegraphics[width = 0.7\textwidth]{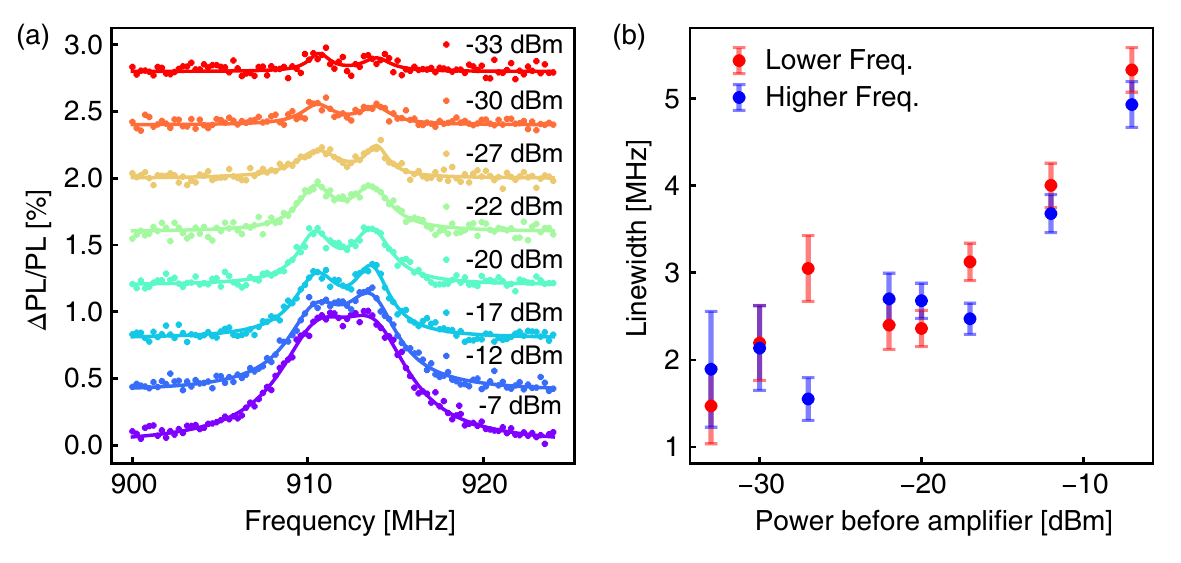}
  \caption{{\bf Microwave power dependence of ODMR spectra.} (a) Continuous-wave ODMR spectra measured at different microwave powers. The plots are offset with each other for comparison. The small splitting is due to ambient stray magnetic field. (b) Power dependence of ODMR linewidths extracted by performing double Lorentzian fits to the spectra in (a).} 
  \label{FigS18}
\end{figure}

We measure the power dependence of continuous-wave ODMR spectra on sample D1. Upon lowering the microwave power, narrower linewidths are observed (\fig{\ref{FigS18}}). When the transition is not power broadened, we measure an inhomogeneous linewidth of $1.47\pm 0.44$~MHz, which corresponds to an ensemble spin dephasing time of $T_2^* = 216 \pm 64$~ns. The spin dephasing time extracted here matches with the result from Ramsey measurement.

\section{\label{SI:Absorption} Absorption measurements}
\begin{figure}[ht]
  \centering
  \includegraphics[width = 0.49\textwidth]{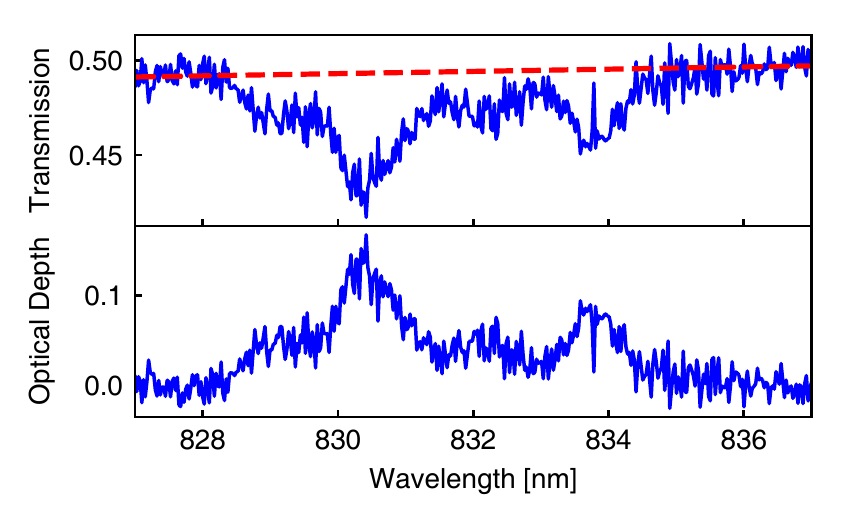}
  \caption{{\bf Absorption measurement data processing.} Upper panel: Transmission ($T$) through the sample normalized by the reference transmission. We fit a straight line as the baseline ($T_s$). Lower panel: Optical depth calculated as $OD = -ln(T/T_s)$. The spike near 834~nm is due to an instability of the laser.} 
  \label{FigS16}
\end{figure}

\section{\label{SI:SaturationCurves} Saturation characteristics of optical spin polarization}
\begin{figure}[h!]
  \centering
  \includegraphics[width = 0.5\textwidth]{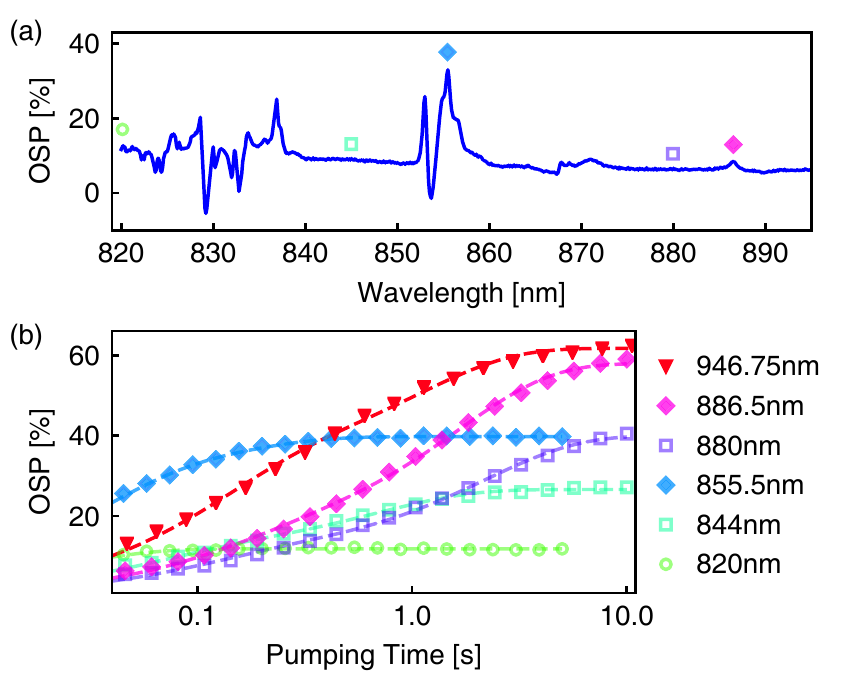}
  \caption{{\bf Saturation of OSP for different wavelengths.} (a) Net OSP measured on sample D2 at 5.5~K. The excitation is kept at constant power (10 mW) and constant pumping time (400 ms). Drastic differences in OSP are observed for BE transitions and off-resonant wavelengths. The markers are visual guidance for the wavelengths used for saturation measurements in (b). Resonant wavelengths are labeled with filled markers while off-resonant wavelengths are labeled with open markers. (b) OSP accumulation as a function of pumping time with 50 mW optical excitation power for all wavelengths. The saturation curves are fitted by a bi-exponential form: $S(t) = a -b_le^{-t/T_l}-(a-b_l)e^{-t/T_s}$.} 
  \label{FigS14}
\end{figure}
We measure the time-dependent saturation characteristics of OSP on sample D2. \fig{\ref{FigS14}(a)} shows an OSP spectrum using constant pumping power and pumping time. The large difference in amplitude between the bound exciton (BE) transitions and off-resonant wavelengths demonstrates the wavelength selectivity of OSP for \siv{} centers. To further characterize the OSP, we measure the time-dependent saturation curves of OSP for several different wavelengths [\fig{\ref{FigS14}(b)}]. The ZPL wavelength (946.75~nm) is included for comparison. The initial spin population is scrambled using off-resonant excitation and a series of MW $\pi/2$ pulses to eliminate any residual polarization from previous interrogation [\fig{\ref{FigS3}(c)}]. Then, optical pulses with varying duration are applied to measure the saturation curve of OSP. The wavelengths are categorized into two groups: off-resonant (820 nm, 844 nm and 880 nm, open markers), and resonant (BE states: 855.5 nm and 886.5 nm and ZPL: 946.75 nm, filled markers). We fit the saturation curves with a bi-exponential function. 

Interpretation of the observed timescales is complicated by the bulk nature of the experiment and the spectrally narrow excitation source, with contributions from far-from-saturation excitation dynamics, spin diffusion, and spin relaxation. However, some qualitative trends are clear; exciting at the ZPL reaches the highest value (62$\%$) but the saturation timescale is rather long. Exciting at the BE transition (855.5~nm), however, shows both high saturation (40$\%$) and a much shorter saturation timescale. OSP saturation
with off-resonant excitation (844~nm and 880~nm) is slow, consistent with our lack of observation of ODMR when detuned from BE transitions [Fig.~4(a) in main text]. For 820 nm excitation above the ionization threshold \cite{Allers1995}, the saturated OSP is small (12$\%$) but the saturation time is fast, likely limited by ionization processes. Strikingly, saturated OSP for 886.5 nm shows a slow timescale but a high saturation value (59$\%$), suggesting high efficiency of OSP per optical cycle.

\section{\label{Sec:Spec_Decomp}Spectral Decomposition of Optical Spin Polarization}
Optical spin polarization is a measure of both absorption and spin polarization from all ground states. The ensemble optical linewidths in bulk samples are much larger than the spin splittings so optical excitation addresses many transitions involving all the spin levels. In order to disentangle the OSP from competing polarization processes, we develop a spectral decomposition method using a pump-probe scheme [\fig{\ref{FigS3}(b)}]. The OSP can be initialized with optical and microwave pulses as 
\begin{equation}
    I_{init}(p_0, p_1, p_{-1}) = p_0 - p_1,
\end{equation}
where $p_0$, $p_1$ and $p_{-1}$ are populations of the three spin sublevels. The population of $m_s = -1$ ($p_{-1}$) is not involved because we are measuring spin echo using $m_s = 0 \leftrightarrow 1$ transition. A weak probe pulse then probes the net change in OSP
\begin{equation}
    I(p_0,p_1,p_{-1}) = \delta p_1(p_0,p_1,p_{-1}) - \delta p_0(p_0,p_1,p_{-1}),
\end{equation}
where $\delta p_i(p_0,p_1,p_{-1})$ is the population change of sublevel $m_s = i$. We assume that the short-time spectrum $I(p_0,p_1,p_{-1})$ will be proportional to the initial population of each spin sublevels ($p_i$) multiplied by their OSP spectra [$I(p_i =1) = \delta p_1(p_i =1) - \delta p_0(p_i =1)$, meaning net OSP change after a perfect $p_i =1$ initialization]. Under this assumption, the net OSP spectrum can be written as a superposition of OSP from all spin sublevels
\begin{equation}
    I(p_0,p_1,p_{-1}) = \sum_{i}p_iI(p_i =1) = \sum_{i}p_i(\delta p_1(p_i =1) - \delta p_0(p_i =1)).
\end{equation}

This expression can be further simplified under specific selection rules as 
\begin{equation}
    I(p_0,p_1,p_{-1}) = \sum_{i}p_i(\delta p_1(p_i =1) - \delta p_0(p_i =1)) = \sum_{i}\Delta_i p_i I_i,
\end{equation}
where we have defined $I_i = \delta p_i (p_i = 1)$ as probe induced population change of sublevel $m_s =i$ after a perfect initialization into sublevel $m_s = i$ and $\Delta_i$ are weight factors depending on the selection rules. 

Here, we consider two generic types of spin selection rules, assuming echo intensity between $m_s= 0$ and $m_s = 1$ is measured. For the first case, we consider no selection rules for optical excitation, meaning that $m_s= 0$ and $m_s = \pm1$ are treated equivalently. An excitation addressing $m_s = -1$ will not lead to any observable effect since the depleted population from $m_s= -1$ is distributed equally to $m_s= 0$ and $m_s = 1$ sublevels. 

For the second case, we consider selection rules similar to magnetic dipole selection rules, where population transfer is only allowed from $m_s= 0$ to $m_s= 1$ and $m_s= 0$ to $m_s= -1$. In this case, when measuring echo intensity between $m_s = 0$ and $m_s = 1$, we expect $\lvert \Delta_1/\Delta_{-1} \rvert = 2$ because the depleted population in $m_s = -1$ is not measured in the echo. We also expect $\lvert \Delta_0/\Delta_{1} \rvert = 0.75$ because half of the depleted population from $m_s = 0$ is transferred to $m_s = -1$ which cannot be measured. The detailed derivation of $\Delta_i$ is summarized in Table~\ref{TabS2}.

\begin{table}[h!]
\caption{\label{TabS2}{Determination of $\Delta_i$ for two types of spin selection rules. We consider the population change of each spin sublevels under spin selective excitation. In each column, we assume perfect initialization into $m_s = i$ ($p_i = 1$), and the depleted population ($\delta$) from $m_s = i$ is redistributed into other sublevels based on the selection rules.}}
\begin{tabular}{lllllllll}
\hline \hline
  \multicolumn{4}{c}{no selection rules} & & \multicolumn{4}{c}{magnetic-dipole-like selection rules} \tabularnewline
\hline 
selective excitation of & $m_s = 0 $ & $m_s = 1$ & $m_s = -1$ & {\hskip 0.5in} & selective excitation of & $m_s = 0$ & $m_s = 1$ & $m_s = -1$\tabularnewline
$\delta p_0$ & $\delta$ & -0.5$\delta$ & -0.5$\delta$ &  & $\delta p_0$ & $\delta$ & -$\delta$ & -$\delta$\tabularnewline
$\delta p_1$ & -0.5$\delta$ & $\delta$ & -0.5$\delta$ &  & $\delta p_1$ & -0.5$\delta$ & $\delta$ & 0\tabularnewline
$\delta p_{-1}$ & -0.5$\delta$ & -0.5$\delta$ & $\delta$ &  & $\delta p_{-1}$ & -0.5$\delta$ & 0 & $\delta$\tabularnewline
$\delta p_1$ - $\delta p_0$  & -1.5$\delta$ & 1.5$\delta$ & 0 &  & $\delta p_1$ - $\delta p_0$ & -1.5$\delta$ & 2$\delta$ & $\delta$\tabularnewline
$\Delta$  & -1.5 & 1.5 & 0 &  & $\Delta$ & -1.5 & 2 & 1\tabularnewline
\hline \hline
\end{tabular}
\end{table}

Because we cannot fully map out the multiplicity and selection rules for these bound exciton states, we choose the first case ($\Delta_0 = -1.5$, $\Delta_1 = 1.5$ and $\Delta_{-1} = 0$) for data processing. The lack of selection rule is a more relaxed requirement since interactions in the excited states could give rise to spin mixing. We note that although choosing a specific set of $\Delta_i$ over another would lead to differences in the relative amplitudes of $I_i$, the resonance features and overall shapes of the spectra are still preserved. 

\begin{figure}[ht!]
  \centering
  \includegraphics[width = 0.5\textwidth]{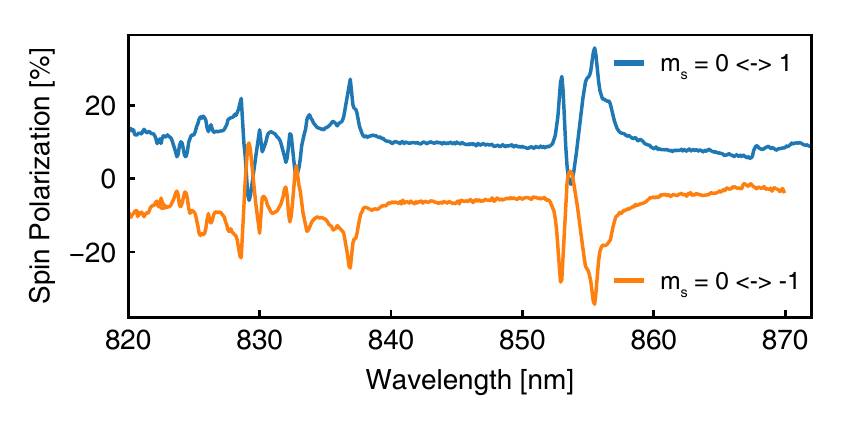}
  \caption{{\bf Comparison of OSP for different spin transitions.} OSP for the two spin transitions ($m_s = 0 \leftrightarrow 1$ and $m_s = 0 \leftrightarrow -1$) are measured on sample D2 with same pumping time (400~ms) and laser power (10~mW). The two spectra are mirror image of each other, consistent with $m_s = 1$ and $m_s = -1$ being symmetric with respect to $m_s = 0$.} 
  \label{FigS5}
\end{figure}

Without any initialization, $p_0 \approx p_1 \approx p_{-1}$, meaning there is equal population in all three spin states. Ideally, the OSP from individual spin levels can be directly measured if $p_i = 1$. In reality, we achieve $p_0 \approx 0.7$ using our most efficient polarization wavelengths. Nevertheless, by initializing the spins differently, individual spectra can be decomposed. When the magnetic field is aligned to the defect axis, $m_s = -1$ and $m_s = 1$ are symmetric with respect to $m_s = 0$ so we could assume $I_{-1} \approx I_1$ and $p_1 \approx p_{-1}$ under $m_s = 0$ initialization. This is consistent with the $m_s = 0 \leftrightarrow 1$ spectrum as the mirror image of $m_s = 0 \leftrightarrow -1$ spectrum, shown in \fig{\ref{FigS5}}. These simplifications lead to
\begin{equation}
    I = \Delta_0p_0I_0 + \Delta_1p_1I_1 + \Delta_{-1} p_{-1}I_{-1} = \Delta_0p_0I_0 + (\Delta_1 + \Delta_{-1})p_1I_1.
\end{equation}

By applying a $\pi$ pulse after the pumping pulse, the spin populations can be inverted
\begin{equation}
    I_{\pi} = \Delta_0p_1I_0 + \Delta_1p_0I_1 + \Delta_{-1} p_{-1}I_{-1} = \Delta_0p_1I_0 + (\Delta_1p_0 + \Delta_{-1} p_{1})I_1.
\end{equation}

The OSP spectra $I_0$ and $I_1$ can then be decomposed from $I$ and $I_\pi$ using the measured initial populations $p_0$ and $p_1$. 



\begin{figure}[h!]
  \centering
  \includegraphics[width = 0.5\textwidth]{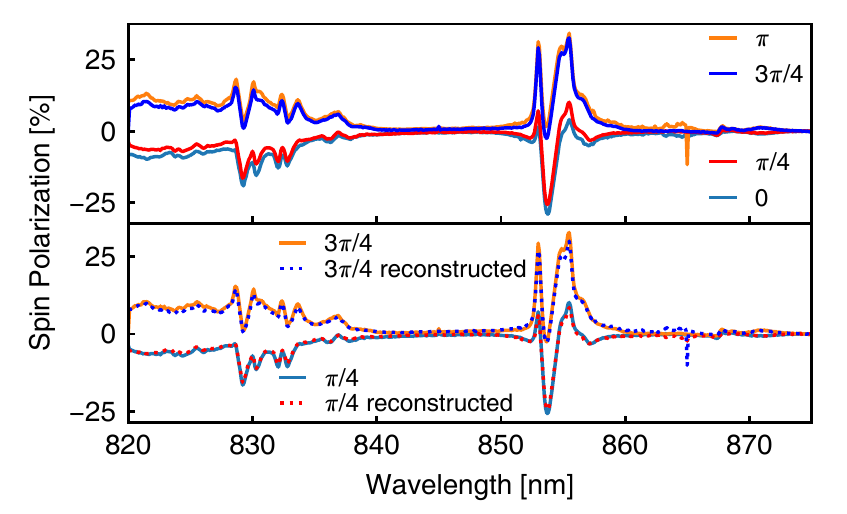}
  \caption{{\bf Spectral decomposition and reconstruction with different MW rotations.} Upper panel: Net OSP with four different MW rotations between pump and probe pulses: 0 (no pulse), $\pi$, $\pi/4$ and $3\pi/4$. Lower panel: Spectral reconstruction of the $\pi/4$ and $3\pi/4$ OSP spectra using the decomposed spectra $I_0$ and $I_1$ from 0 and $\pi$ OSP spectra. The spike near 865 nm is due to an instability of the magnet. The amplitude represents net probe induced OSP change after initialization. This measurement is performed on sample D1. The length of the optical pump pulse (80~mW excitation power) is 4~s, and the length of the optical probe pulse ($\sim$45~mW excitation power) is 100~ms.} 
  \label{FigS6}
\end{figure}

After decomposing the OSP spectrum for different spin states, OSP under arbitrary spin initialization can be reconstructed. To validate the effectiveness of our spectral decomposition, we apply MW pulses with different rotation angles (0, $\pi/4$, $3\pi/4$ and $\pi$) between the pump and probe pulses to achieve different spin initializations. We observe larger net OSP change into $m_s = 0$ ($m_s = 1$) using $\pi$ rotation (0 rotation) compared to $3\pi/4$ rotation ($\pi/4$ rotation), consistent with the difference in the spin initializations (\fig{\ref{FigS6}, upper panel}). Using the decomposed spectra $I_0$ and $I_1$, we could also reconstruct the $\pi/4$ and $3\pi/4$ spectra, which match well with the raw data using $\pi/4$ and $3\pi/4$ rotations (\fig{\ref{FigS6}, lower panel}).

\section{\label{Sec:Isotopes} Isotopic Shifts of the Bound Exciton Transitions}

\begin{figure}[!htbp]
  \centering
  \includegraphics[width = 0.5\textwidth]{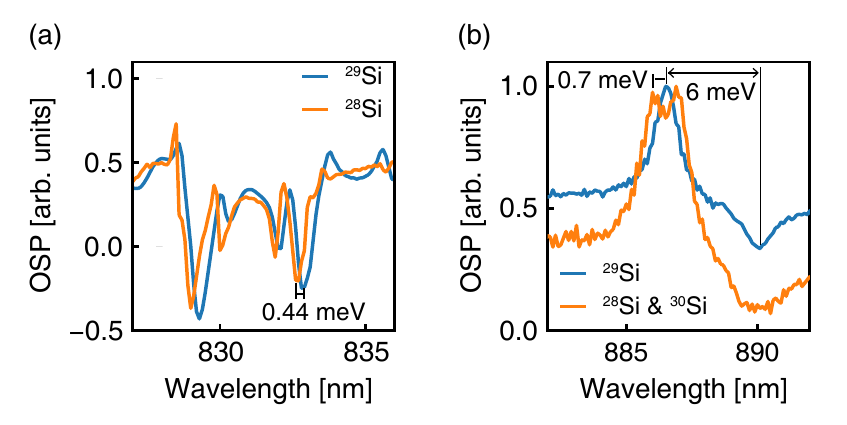}
  \caption{{\bf Isotopic dependence of OSP.}
  (a) Isotopic dependence of $n = 3$ transitions, showing a 0.44~meV shift between $^{28}$Si (orange) and $^{29}$Si (blue). The $^{29}$Si ($^{28}$Si) OSP is measured from sample D1 (D3). (b) Isotopic dependence of the $n = 1$ transitions, showing a 0.7~meV shift between $^{28}$Si (orange) and $^{29}$Si (blue) and between $^{29}$Si (blue) and $^{30}$Si (orange), indicating a different phonon process for $n=1$ transitions. The $^{28}$Si and $^{30}$Si spin transitions are degenerate so they are measured simultaneously on sample D1. The 6 meV splitting between the two transitions is consistent with spin-orbit coupling in the valence band of diamond.} 
 \label{FigS7}
\end{figure}

According to the BE model, the pure electronic transition to the $n = 1$ excited state is dipole forbidden so our level assignment for PLE in Fig.~3(a) starts from $n = 2$. However, we observe OSP resonances near 886~nm that have no correspondence to PLE and absorption peaks [Fig.~2(c)]. These transitions are tentatively assigned to the $n = 1$ transitions. The $n = 1$ states typically do not follow the Rydberg scaling due to the substantial central cell correction expected. Transition to the $n = 1$ state is dipole forbidden so its observation in the OSP spectrum suggests the involvement of a phonon-related process. We find evidence of these phonon processes from the isotopic shift measured on different samples and ESR hyperfine lines. For $n = 3$ transitions, we observe a $\sim$0.4~meV isotopic shift between $^{28}$Si and $^{29}$Si lines [\fig{\ref{FigS7}(a)}], consistent with the isotopic shift observed for the \sivm{} ZPL transition \cite{Dietrich2014}. However, a larger isotopic shift of 0.7~meV is observed for the $n = 1$ transitions [\fig{\ref{FigS7}(b)}] which suggests a different origin of isotopic shift.

\section{\label{sec:SI_theory} THEORETICAL DESCRIPTION OF BOUND EXCITON STATES OF \siv{}}

\subsection{Effective Mass Description}
\label{app:Bohr}
The problem of describing (pseudo-) donor and acceptor defects in the solid state is discussed extensively in many textbooks~\cite{ashcroft1976solid,cardona2005fundamentals}. We revisit some key concepts here to clarify our description in the main text and outline our approach to simulations.

The simplest description of these systems is a hydrogenic model of the pseudo acceptor, where a positive charge is bound to a heavy central negative charge. The Hamiltonian here is thus~\cite{Luttinger1955, Kittel1954, Kohn1955, Wu2008}
\begin{equation}
\hat{H}=-\frac{\hbar^{2}}{2m^{\star}}\Delta-\frac{e^{2}}{4\pi\varepsilon r}\text{,}\label{EqST1}
\end{equation}
where $m^\star$ is the effective mass of the exciton, and $\varepsilon$ is the dielectric constant of the diamond host $\varepsilon=5.7\varepsilon_{0}$. This description neglects the spatial anisotropy imposed by the diamond lattice and the further lowering of symmetry from crystal-field effects introduced by the \siv{} defect. These effects are important and will be discussed below, but this simple model is useful for order-of-magnitude estimates.

The Schr{\"o}dinger equation here can be solved as $\hat{H}\Psi_{n,l,m}=E_{n,l,m}\Psi_{n,l,m}$, where $\Psi_{n,l,m}$ are the hydrogen-atom eigenstates and $E_{n,l,m}$
are their eigenenergies. The energies depend only on the principle quantum number, $n$, so we may write
\begin{equation}
E_{n,l,m}=-\frac{E_{y}}{n^{2}}\qquad E_{y}=\frac{e^{4}m^{\star}}{2(4\pi\varepsilon)^{2}\hbar^{2}}=\frac{m^{\star}}{m_{e}}\frac{\varepsilon^2_0}{\varepsilon^2} E_{Ry} \qquad E_{Ry}=13.6\:\mathrm{eV}\text{.}\label{EqST2}
\end{equation}

The Bohr radius of our artificial atom in diamond can be expressed as
\begin{equation}
r_{n}=r_{0}n^{2}=a_{0}\frac{m_{e}}{m^{\star}}\frac{\varepsilon}{\varepsilon_0}   n^{2}\qquad a_{0}=\frac{4\pi\varepsilon_{0}\hbar^{2}}{m_{e}e^{2}}=0.5292\:\mathrm{\AA}\text{.}\label{EqST3}
\end{equation}

\begin{figure}[h]
  \centering
  \includegraphics[width = 0.40\textwidth]{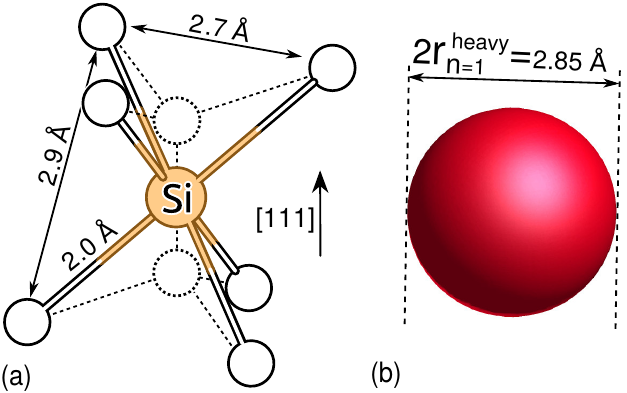}
  \caption{{\bf Relative spatial extent of \siv{} defect and $n=1$ bound exciton.} (a) Geometry of the \siv{} defect. The $e_g$/$e_u$ localized orbitals span the central silicon atom and its six first neighbor carbon atoms. (b) Diameter of the $n=1$ bound exciton state from the Bohr model, see Table~\ref{TabST1}.} 
  \label{FigST1}
\end{figure}

At the $\Gamma$-point in diamond, three different
effective masses are experimentally observed:  $m_{\mathrm{light}}^{\star}=0.7m_{e}$;
$m_{\mathrm{heavy}}^{\star}=2.12m_{e}$; $m_{\mathrm{split-off}}^{\star}=1.06m_{e}$~\cite{1962Rauch, Collins1993optical}. From Table~\ref{TabST1}, we can see that the simplest hydrogenic approximation is poor for all $n=1$ states, and for the heavy hole $n=2$ state. The dimensions of the \siv{} point defect are on the order of a few \AA, comparable to the spatial extent of these wavefunctions as shown in Fig.~\ref{FigST1}.

\begin{table}[h]
\caption{\label{TabST1}The simplest hydrogen atom model for the VBM hole with the corresponding effective masses ($m^{\star}$), the and relative dielectric constant ($\varepsilon$) of diamond as a function of the principal quantum number $n$.}
\begin{tabular}{ccccccc}
n & $E_{n}^{\mathrm{light}}$(eV) & $E_{n}^{\mathrm{heavy}}$(eV) & $E_{n}^{\mathrm{split-off}}$(eV) & $r_{n}^{\mathrm{light}}$(\AA) & $r_{n}^{\mathrm{heavy}}$(\AA) & $r_{n}^{\mathrm{split-off}}$(\AA)\\
\hline 
1 & 0.2930  & 0.8874 & 0.4437 & 4.31 & 1.42 & 2.85\\
2 & 0.0733 & 0.2219 & 0.1109 & 17.24 & 5.69 & 11.38\\
3 & 0.0326 & 0.0986 & 0.0493 & 38.78 & 12.81 & 25.61\\
4 & 0.0183 & 0.0555 & 0.0277 & 68.94 & 22.76 & 45.53\\
$\dots$ & $\dots$ & $\dots$ & $\dots$ & $\dots$ & $\dots$ & $\dots$ \\
$+ \infty$ & 0 & 0 & 0 & $+ \infty$ & $+ \infty$ & $+ \infty$ \\
\end{tabular}
\end{table}

Several features of the experimentally observed transitions are in good agreement with this description. From Fig.~3 of the main text, we extract values of 
\begin{equation}
E_{n}=E_{I}-E_{y}\frac{1}{n^{2}}\qquad E_{I}=1.53\:\mathrm{eV}\qquad E_{y}=0.4\:\mathrm{eV}\text{.}\label{EqST4}
\end{equation}
The ionization threshold is in good agreement with previous photoconductivity measurements, and the effective Rydberg energy is in the range predicted by the light and split-off effective masses (0.293~eV and 0.444~eV, respectively). The data in Fig.~3(b) also shows that the $n=1$ level significantly deviates from the model of a simple hydrogenic series, as expected.

To gain further insight into these states, we go beyond a simple hydrogenic model and explicitly consider the effects of spin-orbit coupling and the crystal field. Parameters which cannot be determined from experimental data in the main text are calculated by DFT, as described in Section~\ref{SI:DFT_All}.


\subsection{Effective Hamiltonian For $n=1$ ($1s$)}

The $\Gamma$-point of the valence band in diamond
is triply degenerate and splits into the light-hole band, heavy-hole band
and split-off band parabolic edges. In order to interpret the character
of the weakly bound hole, we assume that the hole wavefunction is similar
to that of the $\Gamma$-point ($u_{k}(\boldsymbol{r})$).
However, it is confined to the envelope function taken from the
hydrogenic model ($\Psi_{n,l,m}(\boldsymbol{r})$), yielding
\begin{equation}
\Phi(\boldsymbol{r})=u_{k}(\boldsymbol{r})\Psi_{n,l,m}(\boldsymbol{r})\text{.}\label{EqST5}
\end{equation}

For $n=1$, $\Psi_{1,0,0}(\boldsymbol{r})$ is a totally symmetric
$1s$ orbital that transforms as the $A_{1g}$ representation of the local $D_{3d}$
symmetry. The wavefunction at the
$\Gamma$-point in pristine diamond is triply degenerate, which becomes $E_{g}\oplus A_{1g}$ due to the ``crystal-field'' induced by the \siv{} defect.
The $k$ index is $\pm1$ for $E_{g}$ and $k=0$ for the $A_{1g}$ orbital. The total $\Phi(\boldsymbol{r})$ wavefunction transforms
as the product of the two constituent wavefunctions, $A_{1g}\otimes(E_{g}\oplus A_{1g})=E_{g}\oplus A_{1g}$.
In other words, $\Phi(\boldsymbol{r})$ inherits the threefold
multiplicity of the valence band maximum (VBM) states. The following effective Hamiltonian
describes this orbitally three-dimensional hole system,
\begin{equation}
\hat{H}=\hat{H}_{\mathrm{CF}}+\hat{H}_{\mathrm{SO}}=-\frac{\delta}{3}|E_{g}\rangle\langle E_{g}|+\frac{2\delta}{3}|A_{1g}\rangle\langle A_{1g}|+\lambda\left(\hat{L}_{x}\hat{S}_{x}+\hat{L}_{y}\hat{S}_{y}+\hat{L}_{z}\hat{S}_{z}\right)\text{,}\label{EqST6}
\end{equation}
where ``CF'' and ``SO'' are the crystal field and spin-orbit terms, respectively, and $\delta$ and $\lambda$ are the strength of crystal-field and spin-orbit interactions. If we choose the quantization axis along the {[}111{]} direction, parallel with the symmetry axis of the
\siv{} defect, then we may express the operators in Eq.~\eqref{EqST6} as
\begin{equation}
|E_{g}\rangle\langle E_{g}|=\begin{pmatrix}1\\
 & 0\\
 &  & 1
\end{pmatrix}\;|A_{1g}\rangle\langle A_{1g}|=\begin{pmatrix}0\\
 & 1\\
 &  & 0
\end{pmatrix}\quad\hat{L}_{x}=\frac{1}{\sqrt{2}}\begin{pmatrix} & 1\\
1 &  & 1\\
 & 1
\end{pmatrix}\;\hat{L}_{y}=\frac{1}{\sqrt{2}}\begin{pmatrix} & -i\\
+i &  & -i\\
 & +i
\end{pmatrix}\;\hat{L}_{z}=\begin{pmatrix}1\\
 & 0\\
 &  & \!\!-1
\end{pmatrix}\text{.}\label{EqST7}
\end{equation}

The crystal field lifts the degeneracy of the states $|A_{1g}\rangle=(0\;1\;0)$
and $|E_{g+}\rangle=(1\;0\;0)$;
$|E_{g-}\rangle=(0\;0\;1)$ ($|E_{g\pm}\rangle=[|E_{g(x)}\rangle\pm i|E_{g(y)}\rangle]/\sqrt{2}$).
The three orbitals can be treated as an $L=1$ system, where the
$k$ quantum number labels the eigenstates of the $\Gamma$-point as
$\hat{L}_{z}u_{k}(\boldsymbol{r})=k\cdot u_{k}(\boldsymbol{r})$,
or in the matrix representation as $\hat{L}_{z}|A_{1g}\rangle=0|A_{1g}\rangle$ and
$\hat{L}_{z}|E_{g\pm}\rangle=\pm|E_{g\pm}\rangle$. 

Now we consider the effect of the spin-orbit interaction. We assume that the weakly
bound hole is almost spherically symmetric, therefore $\hat{H}_{\mathrm{SO}}$ is also spherically symmetric,
thus it can be described by a single $\lambda$ value. The $\lambda$ parameter can be connected with the spin-orbit splitting of the VBM of diamond: $\Delta_{0}=\frac{3}{2}\lambda$. 

The experimental value of the spin-orbit splitting of diamond
is $\Delta_{0}^{\mathrm{exp.}}=6\:\mathrm{meV}$~\cite{1962Rauch, Herman1963, Serrano1999}. \textit{Ab initio} calculations tend to overestimate this value by a factor of two ($\Delta_{0}\approx13\:\mathrm{meV}$~\cite{Herman1963, Serrano1999, Willatzen1994}), consistent with our
\text{ab initio} DFT calculations yielding $\Delta_{0}^{p.w.}=13.5\:\mathrm{meV}$. We calculated this value on a $\Gamma$-point centered $8\times8\times8$ k-point set for a diamond primitive cell, which results in $\lambda_{\Gamma}^{p.w.}=9.03\:\mathrm{meV}$.
The factor of two between the experimental data and calculated value might indicate the uncertainty in our DFT method or may represent a subtlety in the interpretation of the 6~meV signatures in the spectrum for $\lambda$, as noted in an earlier study~\cite{Serrano1999}. Nevertheless, we cannot unambiguously determine the source of this discrepancy and this issue is beyond the scope of the present manuscript.

We used the following parameters to construct our model directly taken from DFT calculations. According to \textit{ab initio} $\Delta$SCF~\cite{Gali2009PRL} results, the hole experiences $\delta\approx8.8\pm0.1\:\mathrm{meV}$
$D_{3d}$ crystal field  [see Sec.~\ref{sec:2.2} and Fig.~\ref{FigST3}(b)]. The spin-orbit energy is estimated from DFT calculations on the \sivm{} defect and $\lambda\approx9.88\pm0.05\:\mathrm{meV}$ is obtained (Fig.~\ref{FigST5}). The results of the direct diagonalization of this effective Hamiltonian are listed in Table~\ref{TabST2}.

\begin{table}[h!]
\caption{\label{TabST2}Single particle eigenstates of the effective $\hat{H}$
Hamiltonian as given in Eq.~\eqref{EqST6}. Note that the total angular momentum $J$ can be calculated by $J(J+1)=\langle\hat{\boldsymbol{J}}\,\!^{2}\rangle=\langle(\hat{\boldsymbol{L}}+\hat{\boldsymbol{S}})\,\!^{2}\rangle$ formula which can be reduced to the $J=\sqrt{\frac{1}{2}+\ensuremath{\langle\hat{\boldsymbol{L}}+\hat{\boldsymbol{S}}\rangle\,\!^{2}}}-\frac{1}{2}$ expression. Additionally, the presence of the crystal field splitting ($\delta$) reduces rotational symmetry, thus $J$ is not a good number and its expectation value is not a half integer. Nevertheless, without any $D_{3d}$ crystal field $J$ is a good quantum number. The heavy and light-hole bands combine to form a 4-fold degenerate $J=L+S=3/2$ level, and the split-off band constitutes the $J=L-S=1/2$ level. In other words, if one considers the effect of the crystal field, which lowers the spherical symmetry to $D_{3d}$ point group symmetry, then the two $E_{\frac{1}{2}g}$ representations are allowed to mix with each other and, as a consequence, the $J$ quantum number deviates from the half-integer value. }
\begin{ruledtabular}
\begin{tabular}{clcccccccc}
 &  & energy (meV) & L & S & J & $\langle\hat{J}_{z}\rangle$ & $\langle\hat{L}_{z}\rangle$ & $\langle\hat{S}_{z}\rangle$ & $\langle\hat{\boldsymbol{L}}\hat{\boldsymbol{S}}\rangle$\tabularnewline
\hline 
\multirow{2}{*}{$E_{\frac{1}{2}g}$} & \multirow{2}{*}{split-off band} & -10.802 & 1 & 0.5 & 0.57 & 0.5 & 0.85 & -0.35 & -0.43\tabularnewline
 &  & -10.802 & 1 & 0.5 & 0.57 & -0.5 & -0.85 & 0.35 & -0.43\tabularnewline
\hline 
\multirow{2}{*}{$E_{\frac{3}{2}g}$} & \multirow{2}{*}{heavy-hole band} & 2.007 & 1 & 0.5 & 1.50 & 1.5 & 1 & 0.50 & 0.50\tabularnewline
 &  & 2.007 & 1 & 0.5 & 1.50 & -1.5 & -1 & -0.50 & 0.50\tabularnewline
\hline 
\multirow{2}{*}{$E_{\frac{1}{2}g}$} & \multirow{2}{*}{light-hole band} & 8.795 & 1 & 0.5 & 1.46 & 0.5 & 0.15 & 0.35 & -0.07\tabularnewline
 &  & 8.795 & 1 & 0.5 & 1.46 & -0.5 & -0.15 & -0.35 & -0.07\tabularnewline
\end{tabular}
\end{ruledtabular}
\end{table}

We calculate a $\sim$9~meV splitting between the quasi-particle hole levels.
The experimentally-observed splitting between the two $n=1$ spin polarization resonances is 5.86~meV (difference between 886~nm and 889~nm peaks in the spectrum), which is consistent with the experimental value of the spin-orbit parameter. The discrepancy between our calculation and experimentally observed values is consistent with the general observation that \textit{ab initio} methods appear to overestimate this value.

We have so far only considered the wavefunction of the hole. The resulting \sivm{} defect also has non-zero spin, and can be described by a second hole localized on the defect. This second hole may be described as a $^{2}E_{g}$ orbitally degenerate spin-half system that splits into the $E_{\frac{1}{2}g}\oplus E_{\frac{3}{2}g}$ Kramers doublets. Under the assumption that the two holes are independent of each other, we can construct the two-hole wavefunction as a direct product of the localized hole in the \sivm{} $e_g$ orbital and the weakly-bound hole as follows:  
\begin{equation}
\mathrm{split-off\:hole}:\quad E_{\frac{1}{2}g}\otimes\left(E_{\frac{1}{2}g}\oplus E_{\frac{3}{2}g}\right)=A_{1g}\oplus A_{2g}\oplus3E_{g}\text{;}\label{EqST8}
\end{equation}

\begin{equation}
\mathrm{heavy\:hole:}\quad E_{\frac{3}{2}g}\otimes\left(E_{\frac{1}{2}g}\oplus E_{\frac{3}{2}g}\right)=2A_{1g}\oplus2A_{2g}\oplus2E_{g}\text{;}\label{EqST9}
\end{equation}

\begin{equation}
\mathrm{light\:hole}:\quad E_{\frac{1}{2}g}\otimes\left(E_{\frac{1}{2}g}\oplus E_{\frac{3}{2}g}\right)=A_{1g}\oplus A_{2g}\oplus3E_{g}\text{.}\label{EqST10}
\end{equation}

One finds from Eqs.~(\ref{EqST8}-\ref{EqST10}) that the $E_{\frac{1}{2}g}$ split-off hole, the $E_{\frac{3}{2}g}$ heavy hole, and the $ E_{\frac{1}{2}g}$ light hole Kramers doublets will split further due to the coupling of the additional $e_g$ hole. Note, the $A_{1g}, A_{2g}, E_g$ representations are double group representations. 

\begin{table}[h!]
\caption{\label{TabST3}Two-hole eigenstates of the effective $\hat{H}$ Hamiltonian of Eq.~\eqref{EqST11}. We depict the relative contributions of the $S=0$ singlet and $S=1$ triplet subspaces for each eigenstate. We also list the $m_s=0$ and $m_s=\pm1$ projections inside the triplet subspace. For example, one can say that the heavy hole with 2.007~meV energy is 100\% composed of an $m_s=\pm1$ triplet. However, the light hole with 8.506~meV exhibits only 85\% singlet probability since the spin-orbit coupling introduces 15\% $m_s=\pm1$ character into this dominantly singlet state.}
\begin{tabular}{lcllll}
\hline \hline
 &  &  &  & \multicolumn{2}{c}{triplets}\tabularnewline
 &  &  & singlets & \multicolumn{2}{c}{$\swarrow S=1\searrow$}\tabularnewline
 &  & energy (meV) & $S=0$ & $m_{s}=0$ & $m_{s}=\pm1$\tabularnewline
\hline 
\multirow{8}{*} & \multirow{8}{*}{split-off band} & -10.948 & 0.44 & 0.42 & 0.15\tabularnewline
 &  & -10.948 & 0.44 & 0.42 & 0.15\tabularnewline
 &  & -10.948 & 0.44 & 0.42 & 0.15\tabularnewline
 &  & -10.948 & 0.44 & 0.42 & 0.15\tabularnewline
 &  & -10.853 & 0.15 & 0 & 0.85\tabularnewline
 &  & -10.853 & 0.15 & 0 & 0.85\tabularnewline
 &  & -10.802 & 0 & 0.15 & 0.85\tabularnewline
 &  & -10.802 & 0 & 0.15 & 0.85\tabularnewline
\hline 
\multirow{8}{*} & \multirow{8}{*}{heavy-hole band} & 1.838 & 0.49 & 0.51 & 0\tabularnewline
 &  & 1.838 & 0.49 & 0.51 & 0\tabularnewline
 &  & 1.838 & 0.49 & 0.51 & 0\tabularnewline
 &  & 1.838 & 0.49 & 0.51 & 0\tabularnewline
 &  & 2.007 & 0 & 0 & 1\tabularnewline
 &  & 2.007 & 0 & 0 & 1\tabularnewline
 &  & 2.007 & 0 & 0 & 1\tabularnewline
 &  & 2.007 & 0 & 0 & 1\tabularnewline
\hline 
\multirow{8}{*} & \multirow{8}{*}{light-hole band} & 8.506 & 0.85 & 0 & 0.15\tabularnewline
 &  & 8.506 & 0.85 & 0 & 0.15\tabularnewline
 &  & 8.770 & 0.07 & 0.08 & 0.85\tabularnewline
 &  & 8.770 & 0.07 & 0.08 & 0.85\tabularnewline
 &  & 8.770 & 0.07 & 0.08 & 0.85\tabularnewline
 &  & 8.770 & 0.07 & 0.08 & 0.85\tabularnewline
 &  & 8.795 & 0 & 0.85 & 0.15\tabularnewline
 &  & 8.795 & 0 & 0.85 & 0.15\tabularnewline
\hline \hline
\end{tabular}
\end{table}

We note that both singlet and triplet spin configurations appear in these two-hole wavefunctions. The energy levels of these states cannot be predicted by this simple model and we approximate those by \textit{ab initio} simulations. We again rely on the $\Delta$SCF method. We can impose a triplet
coupling between the localized $e_{g}$ hole and the weakly bound hole, thus their spin state is maximally polarized $\left|\uparrow\uparrow\right\rangle$. However, we can
also determine when the two holes exhibit different spin projections such as $\left|\uparrow\downarrow\right\rangle$ which mimics the singlet configuration.

\begin{figure}[h!]
\includegraphics[width=0.9\columnwidth]{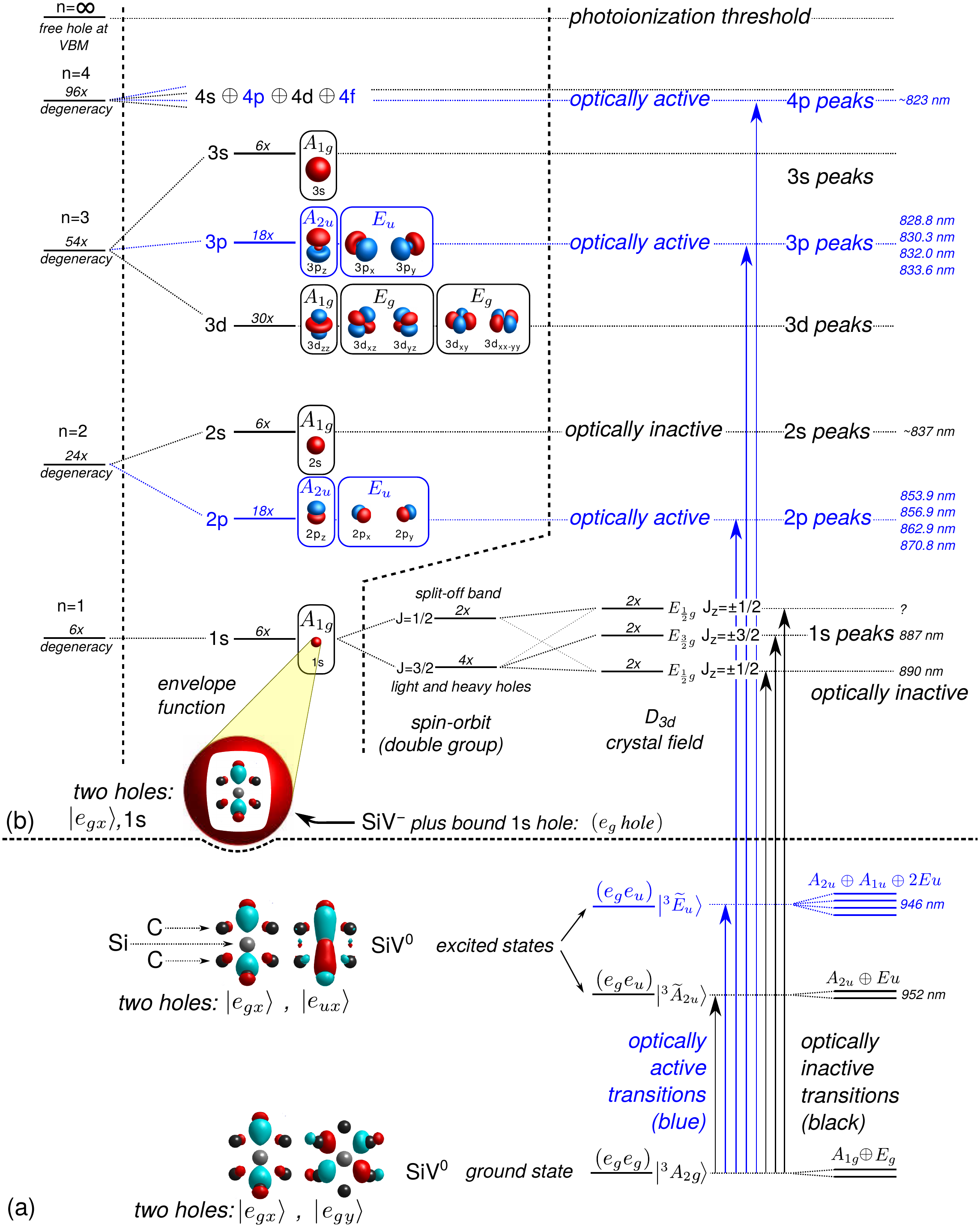}
\caption{{\bf Schematic overview of the bound exciton model.} (a) Depiction of orbitals and energy levels associated with ground and excited states of \siv{}. (b) Depictions of the bound exciton excitations, where the model is a \sivm{} with a weakly bound hole that is orbiting around the negative defect.}
\label{FigST2}
\end{figure}

In this way, we are able to determine the difference between the singlet and triplet
states as $\Lambda=0.34\pm0.01$~meV, see Sec.~\ref{sec:tripletsinglet} for details. Our DFT results follow Hund's rule in that the triplet configuration is lower in energy than the singlet configuration. Thus our effective Hamiltonian now becomes
\begin{equation}
\hat{H}=\hat{H}_{\mathrm{CF}}+\hat{H}_{\mathrm{SO}}+\Lambda|\mathrm{singlets}\rangle\langle\mathrm{singlets}|\text{,}\label{EqST11}
\end{equation}
where the singlet operator raises the energy of the singlet states as  $|\mathrm{singlets}\rangle=(\left|\uparrow\downarrow\right\rangle-\left|\downarrow\uparrow\right\rangle)/\sqrt{2}$ while leaving the three triplet projections $m_s=-1,0,+1$ untouched. Considering the two-hole wavefunction significantly increases the dimensionality of the problem, as can be seen from Table~\ref{TabST3}. However, the coupling of the second hole only perturbatively splits the three levels by $\sim$0.2~meV. Thus the single hole picture from Table~\ref{TabST2} is representative of the physical nature of the system.

\subsection{Extension to the $n>1$ Bound Excitons}

The computational complexity of the system increases rapidly with $n$, thus our calculations for the $n=2$ transitions in Sec.~\ref{Sec:n2RydbergDFT} are only a crude approximation. However, we can use the physical intuition we developed for the $n=1$ states to describe some properties of these states. We summarize the energy levels of the bound exciton states in Fig.~\ref{FigST2}.

For example, at $n=2$, four different envelope functions are possible with $\Psi_{2,l,m}(\boldsymbol{r})$.
There is a ``2s'' hole with $l=0$ and $m=0$ that transforms as
$A_{1g}$. There is also a 3-fold degenerate ``2p'' solution, which under the $D_{3d}$ crystal field splits into ${2p_{z}}$
with $A_{2u}$ representation and (${2p_{x}}$,${2p_{y}}$) with $E_{u}$ representation. These states are visually depicted in Fig.~\ref{FigST2}.

We make the following observations regarding the $n=2,3$ manifolds:
\begin{itemize}
\item If the primary source of the spin-orbit splitting again comes from the
$u_{k}(\boldsymbol{r})$ wavefunction at the $\Gamma$-point then the spin-orbit interaction for $n=2,3$ levels will be similar to the $n=1$ value, approximately $\lambda\sim10$~meV.
In this case, this splitting would be independent of $n$ since the $u_{k}(\boldsymbol{r})$
wavefunction will be the same for all $n=1\dots\infty$.
\item Only transitions to states with $p$-like envelope functions are expected to be optically active. Clusters of four peaks are observed experimentally for ``2p'' and ``3p'', see Fig.~3(a) of the main text.  
\item The energies of ``1s'', ``2s'' peaks deviate most significantly from the Bohr model Eq.~\eqref{EqST4} [Fig.~3(b)]. However the ``2p'' and ``3p'' states largely follow the $1/n^2$ law of the Bohr model. We treat this difference as a central cell correction which alters the energy level of the ``1s'' state significantly (see Sec.~\ref{Sec:n2RydbergDFT} for details). The localized $e_g$ orbital of the \sivm{} excludes the $1s$ from the 6 first neighbor carbon atoms, where it would exhibit otherwise the highest probability density, increasing the spatial extent of this wavefunction. This central cell correction effectively increases the excitation energy of ``1s'' and ``2s'', but leaves ``2p'' and ``3p'' intact due to the radial node at the origin in their probability density.
\item It is extremely complicated to setup an effective Hamiltonian for ``2p'' and ``3p'' states in a similar fashion as we did for ``1s'' in Eq.~\eqref{EqST6}. Not only would the $u_{k}(\boldsymbol{r})$ wavefunction at the $\Gamma$-point would carry the orbital momentum of $L_z=-1,0,+1$, but the envelope function of ``p'' orbitals would also exhibit an $L=1$ angular momentum. However, the ``2s'' should behave very similarly to ``1s'', albeit with altered crystal-field and spin-orbit parameters.
\end{itemize}

\section{\label{SI:DFT_All} Results of DFT calculations for the energy levels of Bound Exciton resonances} 
\subsection{Method Summary}
First principles plane-wave supercell DFT calculations are used to study the \siv{} center in diamond as implemented in the \textsc{vasp} code~\cite{Kresse:PRB1996}. The excited states are considered by the $\Delta$SCF method which involves electron-hole interaction and relaxation of ions upon excitation~\cite{Gali2009PRL}. The paramagnetic states are treated by spin polarized functionals. The spin-orbit energies are calculated within the scalar relativistic approximation~\cite{Steiner2016}. The usual projector augmented wave (PAW) projectors~\cite{Blochl:PRB1994,Blochl:PRB2000} are applied on the carbon and silicon atoms with a plane wave cutoff of 420~eV. We provide a foundation for the accurate calculation of the effective mass (acceptor) states within supercell modeling in the subsequent sections. Our approach requires scaling of the properties as a function of supercell size. In the scaling procedure, supercells of up to 8000 atoms are applied within the semilocal Perdew-Burke-Ernzerhof (PBE) DFT functional~\cite{PBE}, whereas supercells of up to 1000 atoms are employed in the hybrid Heyd-Scuseria-Ernzerhof (HSE) DFT functional~\cite{Heyd03, Krukau06}.

\subsection{Determining the $n=1$, $\delta$ Parameter From Kohn-Sham
Levels\label{sec:2.1}}

We calculate the electronic structure of an \sivm{} defect embedded in diamond cubic supercells of 64, 216, 512, 1000, 1728, 2744, 4096, 5832, and 8000 carbon
atoms within $\Gamma$-point sampling of the Brillouin-zone without incorporating the spin-orbit interaction by means of the PBE~\cite{PBE} DFT functional. In this case, the VBM at the $\Gamma$-point should be triply-degenerate in the perfect supercell calculation. However, due
to the presence of the crystal field induced by the defect in the defective supercell, the cubic symmetry of the supercell is lowered to $D_{3d}$, thus the VBM at the $\Gamma$-point splits into $e_{g}$ and $a_{1g}$ states. We define the Kohn-Sham energy
difference of these two as $\delta=E(a_{1g})-E(e_{g})$. The optical
excitation process can be described as promotion of an electron from the delocalized $a_{1g}$ or $e_{g}$ levels to the unoccupied and localized $e_{g}$ level
in the same spin channel. We use spin majority (minority) channel to refer to the appropriate Kohn-Sham level in the calculation. As an example, for a $S=1$ spin system in the $m_s=+1$
configuration, the spin-up electrons are in the majority, whereas in the $m_s=-1$
configuration the situation is reversed.

\begin{figure}
\includegraphics[scale=0.60]{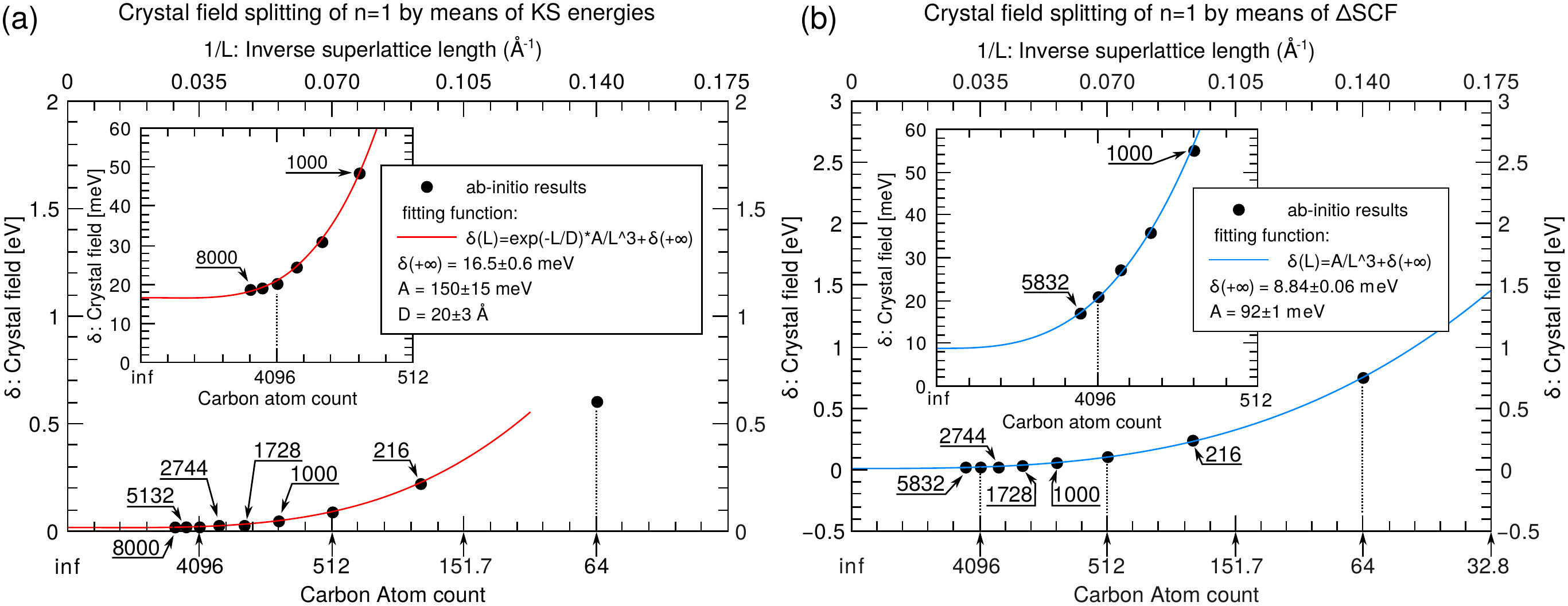}\caption{
\label{FigST3}{Crystal field splitting. We fit the $\delta_\mathrm{KS}$ and $\delta_\mathrm{tot}$ energy from Table~\ref{TS4} versus supercell size.}}
\end{figure}

Table~\ref{TS4} lists the Kohn-Sham
energies of the VB states and the localized $e_{g}$ and $e^{\prime}_{u}$ orbitals for the excitation process of \siv{} that occurs at 946~nm (taking the relaxation of ions upon excitation into account). We note
that the in-gap localized $e^{\prime}_{g}$ level is occupied only by one electron, so we put half-half electrons onto $e^{\prime}_{g(x)}$ and $e^{\prime}_{g(y)}$ states, in order to average out the Jahn-Teller instability of $^{2}E_{g}$ state of \sivm.

\begin{table}
\caption{
\label{TS4}{Kohn-Sham eigenstates of 64-8000
atom supercells in the spin minority channel. This spin channel corresponds to 
the spin allowed optical transitions that are triplet-to-triplet transitions of \siv.}
}
\begin{tabular}{llllllllllll}
\hline \hline
C atom count &  & 64 & 216 & 512 & 1000 & 1728 & 2744 & 4096 & 5832 & 8000 & $+\infty$\tabularnewline
lattice constant & \AA & 7.13 & 10.70 & 14.26 & 17.84 & 21.40 & 24.97 & 28.54 & 32.10 & 35.67 & $+\infty$ \tabularnewline
\hline 
\multirow{2}{*}{localized $\varepsilon_\mathrm{KS}(e_{u}^{\prime\uparrow})$} & eV & 8.810 & 9.389 & 9.308 & 9.432 & 9.516 & 9.575 & 9.616 & 9.646 & 9.668 & \tabularnewline
 & eV & 8.810 & 9.389 & 9.308 & 9.432 & 9.516 & 9.575 & 9.616 & 9.646 & 9.668 & \tabularnewline
\multirow{2}{*}{delocalized $\varepsilon_\mathrm{KS}(e_{g}^{\uparrow})$} & eV & 8.894 & 9.710 & 9.615 & 9.675 & 9.702 & 9.716 & 9.725 & 9.730 & 9.735 & \tabularnewline
 & eV & 8.894 & 9.710 & 9.616 & 9.675 & 9.702 & 9.716 & 9.725 & 9.730 & 9.735 & \tabularnewline delocalized
$\varepsilon_\mathrm{KS}(a_{1g}^{\uparrow})$ & eV & 9.500 & 9.931 & 9.705 & 9.724 & 9.733 & 9.740 & 9.745 & 9.749 & 9.754 & \tabularnewline
\multirow{2}{*}{localized $\varepsilon_\mathrm{KS}(e_{g}^{\prime\uparrow})$} & eV & 11.082 & 11.090 & 10.831 & 10.886 & 10.934 & 10.975 & 11.007 & 11.033 & 11.055 & \tabularnewline
 & eV & 11.082 & 11.090 & 10.831 & 10.886 & 10.934 & 10.975 & 11.007 & 11.033 & 11.055 & \tabularnewline
 \hline 
$\delta_\mathrm{KS}=\varepsilon_\mathrm{KS}(a_{1g}^{\uparrow})-\varepsilon_\mathrm{KS}(e_{g}^{\uparrow})$ & meV & 605.3 & 221.5 & 89.7 & 48.6 & 31.1 & 24.5 & 20.3 & 19.1 & 18.8 & 18.5$\pm$0.2\tabularnewline
$\delta_\mathrm{tot}=E_\mathrm{tot}(a_{1g}^{\uparrow})-E_\mathrm{tot}(e_{g}^{\uparrow})$ & meV & 757.6 & 247.4 & 102.4 & 55.0 & 35.9 & 27.1 & 20.8 & 18.5 &    & 7.91 \tabularnewline
\hline \hline
\end{tabular} 
\end{table}

Fig.~\ref{FigST3}(a) shows the $\delta$ crystal-field parameter as calculated from the energy gap of the Kohn-Sham levels. In order to scale the result to the infinite system (isolated defect), we fit an $\delta(L)=\frac{A}{L^3}\exp(-\frac{L}{D})+\delta(+\infty)$ function to the data ranging in size from 216-atom to 8000-atom supercells. The Kohn-Sham energies, however, are auxiliary quantities in Kohn-Sham DFT, thus we move to the next task of calculating the total energy differences by means of $\Delta$SCF method.

\subsection{$n=1$, $\delta$ Parameter From $\Delta$SCF
calculations\label{sec:2.2}}

To take into account the electron-hole interaction, we calculate the total energies by $\Delta$SCF method at the PBE level, where we leave and constrain a hole inside a VBM state, and then converge the electronic structure with this constraint. We calculate the total energy of the \sivm{} plus a hole left behind in the delocalized $a_{1g}$ state, and also where the hole left behind is in the delocalized $e_{g}$ state. The calculated $\Delta$SCF energies are scaled by a fit function $\delta(L)=\frac{A}{L^3}+\delta(+\infty)$ [\fig{\ref{FigST3}(b)}]. The fit describes the crystal field from total energy differences [$\delta_{\mathrm{tot}}=E_{\mathrm{tot}}(a_{1g})-E_{\mathrm{tot}}(e_{g})$] as a function of supercell sizes ranging from 64-atom to 8000-atom supercells. The fit yields $\delta\approx8.84\pm0.06$~meV.

\subsection{Spin-Orbit Coupling at the $\Gamma$-Point\label{sec:2.3}}

\begin{figure}[]
\includegraphics[scale=0.7]{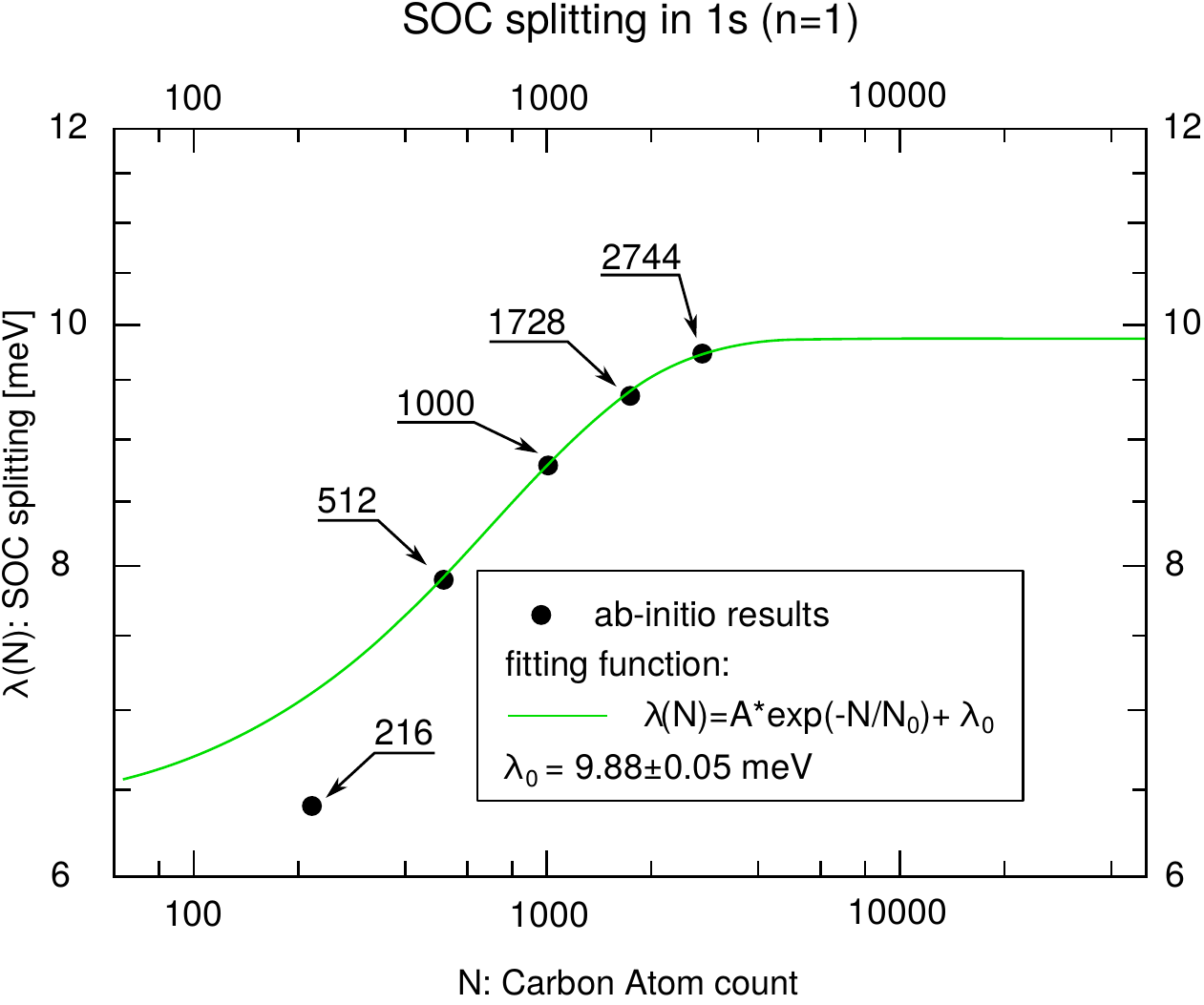}\caption{\label{FigST5}{Supercell scaling of spin-orbit coupling.}}
\end{figure}

Calculations of the spin-orbit coupling are performed as described
previously~\cite{Thiering2018PRX}. We calculate the ground
state of \sivm{} and determine the spin-orbit splitting of the
$e_{g}$ delocalized level in the $\Gamma$-point at the VBM. We can
determine the spin-orbit splitting $\lambda$ as the energy difference
between $e_{g+}^{\uparrow}$ and $e_{g-}^{\uparrow}$ levels. We find that the accurate calculation of this property requires scaling of supercell sizes as shown in Fig.~\ref{FigST5}, where we fit an exponential scaling function to achieve the isolated defect limit with an infinitely large $L\rightarrow + \infty$ supercell. The 216-atom supercell is too small for this quantity within $\Gamma$-point approximation and is not taken into account in the fitting procedure.

\subsection{Triplet-Singlet Splitting of the $n=1$ Series}
\label{sec:tripletsinglet}
We determined the strength of triplet-singlet splitting from the single particle Kohn-Sham levels of \sivm{} center. Table~\ref{TS4} shows the spin minority ($\uparrow$) channel, where the optical transition occurs (one electron fills the double degenerate $e_g^{\prime\uparrow}$ in the ground state which is fully occupied in the excited state). Table~\ref{TabST4} lists the Kohn-Sham levels in the spin majority channel, where the $e_g^{\prime\downarrow}$ is fully occupied by two electrons. While the transitions from any occupied single electron orbital from the spin minority channel are spin allowed upon optical excitation, the excitation process which flips the spin is forbidden (because of the relatively small spin-orbit interaction). Therefore the energy splitting $\Lambda=\varepsilon(a_{1g}^{\downarrow})-\varepsilon(a_{1g}^{\uparrow})$ for the same orbitals but with the opposite spin channel provides insight into the spin forbidden transition. That is, $\Lambda$ gives a tentative approximation for the energy difference when the ``1s'' $a_{1g}$ hole couples with the $e^\prime_g$ with a spin triplet or spin singlet wavefunction. We note that the ``1s'' $e_{g}$ exhibits different triplet-singlet splitting ($\lambda_e$). However, we use only the definition of $\Lambda$ to derive the full singlet manifold, see Eq.~\eqref{EqST11}. 

\begin{table}[]
\caption{\label{TabST4}{Kohn-Sham levels in the spin majority channel as obtained in the 64-8000 atom supercells. This spin channel corresponds to the
the spin forbidden optical transitions. These are the triplet to singlet
transitions, where the electron spin is flipped. The definition of triplet-singlet splitting ($\Lambda$) for the $a_{1g}$ ``1s'' involves the values from the spin up channel from Table~\ref{TS4}.}}
\begin{tabular}{llllllllllll}
\hline \hline
C atom count &  & 64 & 216 & 512 & 1000 & 1728 & 2744 & 4096 & 5832 & 8000 & $+\infty$ \tabularnewline
\hline 
\multirow{2}{*}{$\varepsilon(e_{u}^{\prime\downarrow})$-localized} & eV & 8.726 & 9.261 & 9.219 & 9.377 & 9.482 & 9.553 & 9.601 & 9.634 & 9.659 & \tabularnewline
 & eV & 8.726 & 9.261 & 9.219 & 9.377 & 9.482 & 9.553 & 9.601 & 9.634 & 9.659 & \tabularnewline
\multirow{2}{*}{$\varepsilon(e_{g}^{\downarrow})$-delocalized} & eV & 8.731 & 9.671 & 9.599 & 9.667 & 9.697 & 9.712 & 9.722 & 9.728 & 9.733 & \tabularnewline
 & eV & 8.731 & 9.671 & 9.599 & 9.667 & 9.697 & 9.712 & 9.722 & 9.728 & 9.733 & \tabularnewline
$\varepsilon(a_{1g}^{\downarrow})$-delocalized & eV & 9.490 & 9.928 & 9.704 & 9.723 & 9.732 & 9.740 & 9.744 & 9.749 & 9.753 & \tabularnewline
\multirow{2}{*}{$\varepsilon(e_{g}^{\prime\downarrow})$-localized} & eV & 10.884 & 10.880 & 10.599 & 10.644 & 10.689 & 10.729 & 10.760 & 10.785 & 10.807 & \tabularnewline
 & eV & 10.884 & 10.880 & 10.599 & 10.644 & 10.689 & 10.729 & 10.760 & 10.785 & 10.807 & \tabularnewline
\hline 
$\delta_\mathrm{KS}^{\prime}=\varepsilon(a_{1g}^{\downarrow})-\varepsilon(e_{g}^{\downarrow})$ & meV & 758.7 & 257.4 & 105.0 & 56.0 & 35.4 & 27.4 & 22.4 & 20.8 & 18.2 &18.5$\pm$0.2\tabularnewline
$\Lambda=\varepsilon(a_{1g}^{\downarrow})-\varepsilon(a_{1g}^{\uparrow})$ & meV & 9.92 & 3.45  & 1.65 & 0.99  & 0.71  & 0.59  & 0.50 & 0.53  & 0.54  & 0.34$\pm$0.01 \tabularnewline
$\Lambda_{e}=\varepsilon(e_{g}^{\downarrow})-\varepsilon(e_{g}^{\uparrow})$ & meV & 163.36  & 39.31  & 16.86  & 8.37  & 4.97  & 3.40  & 2.60  & 2.19  & 1.97  & 0.49$\pm$0.13\tabularnewline
\hline \hline
\end{tabular}
\end{table}

\subsection{Determining the $n=1$ Excitation Energy by Means of $\Delta$SCF Calculation
at the HSE06 Level}
\label{sec:HSE06.deltaSCF}

Although the PBE functional provides insight into the nature of the bound exciton states, it underestimates the optical excitation energies. To enable quantitative comparison to experimental data, we determine the optical excitation energies by means of the HSE06 functional~\cite{Heyd03, Krukau06}, which provides approximately $0.1$~eV accuracy for the excitation and ionization energies of point defects in diamond. However, supercells containing more than 1000 carbon atoms are computationally too expensive using the HSE06 functional. Thus, we exploit the two functionals for two different purposes: the PBE functional is used to simulate the supercell size dependence of these properties at the $N\rightarrow \infty$ limit, i.e., completely isolated defect limit, and the HSE06 functional is used to correct the optical excitation energies by comparing the PBE and HSE06 results at smaller supercells. We summarize the optical excitation processes in Fig.~\ref{FigST6} and Table~\ref{TabST5}, where we depict three excitation processes:
\begin{itemize}
\item Excitation from the localized $e_g$ orbital into $e_u$: $\Delta E_{\mathrm{loc}}$. This is the optical transition reported previously~\cite{Rose2018,Green2017,Green2019}
\item Excitation from the localized $e_g$ orbital into the $n=1$ bound exciton state: $\Delta E_{n=1}$.
\item Excitation from the localized $e_g$ orbital into the $n=\infty$ bound exciton state: $\Delta E_{n=\infty}$. This is the acceptor adiabatic charge transition level.
\end{itemize}

\begin{table}[h!]
\caption{\label{TabST5}{\textit{Ab initio} results and their comparison to the experimental data}}
\begin{tabular}{lllllllll}
\hline \hline
 &  & \multicolumn{2}{l}{$|e_{g}\rangle\leftrightarrow|e_{u}\rangle$ } & \multicolumn{3}{l}{$n=1$ BE excitation} & \multicolumn{2}{c}{charge transition level ($n=\infty$)}\tabularnewline
$N$ & $1/L\:(\mathrm{\AA})$  & $\Delta E_{\mathrm{loc}}^{\mathrm{HSE06}}$ & $\Delta E_{\mathrm{loc}}^{\mathrm{corrected}}$ & $\Delta E_{n=1}^{\mathrm{PBE}}$ & $\Delta E_{n=1}^{\mathrm{HSE06}}$ & $\Delta E_{n=1}^{\mathrm{corrected}}$ & $\Delta E_{n\rightarrow\infty}^{\mathrm{PBE}}$ & $\Delta E_{n\rightarrow\infty}^{\mathrm{HSE06}}$\tabularnewline
\hline 
64 & 0.140 & 1.972 & \textbf{1.943} & 1.445 & 1.573  & \textbf{1.616} & 1.293 & \textbf{1.741}\tabularnewline
216 & 0.093 & 1.489 & \textbf{1.459} & 1.019 & 1.234 & \textbf{1.276} & 0.975 & \textbf{1.309}\tabularnewline
512 & 0.070 & 1.374 & \textbf{1.344} & 0.946 & 1.224 & \textbf{1.267} & 0.929 & \textbf{1.259}\tabularnewline
1000 & 0.056 & 1.328 & \textbf{1.298} & 0.943 & 1.256 & \textbf{1.298} & 0.937 & \textbf{1.276}\tabularnewline
1728 & 0.047 &  &  & 0.953 &  &  & 0.952 & \tabularnewline
2744 & 0.040 &  &  & 0.966 &  &  & 0.971 & \tabularnewline
4096 & 0.035 &  &  & 0.977 &  &  & 0.985 & \tabularnewline
5832 & 0.031 &  &  & 0.986 &  &  & 0.997 & \tabularnewline
8000 & 0.028 &  &  &  &  &  &  & \tabularnewline
$+\infty$ & 0 & 1.297 & \textbf{1.268} & 1.034 & 1.437 & \textbf{1.480} & 1.121 & \textbf{1.547}\tabularnewline
exp. &  &  & \textbf{1.311}  &  &  & \textbf{1.393} &  & \textbf{1.53}\tabularnewline
\hline \hline
\end{tabular}
\end{table}

\subsubsection{Excitation between the localized $e_g \leftrightarrow e_u$ states}
We determined the excitation process in various supercell sizes between $N=64\dots1000$ carbon atoms by means of HSE06 functional. The $\Delta E_{\mathrm{loc}}$ excitation process that corresponds to the zero-phonon line optical signals at 946 and 951~nm can be expressed by the following formula:
\begin{equation}
\Delta E_{\mathrm{loc}}^{\mathrm{HSE06}}(L)=E_{\mathrm{tot}}[\mathrm{SiV}{}_{\mathrm{excited}}^{0}](L)-E_{\mathrm{tot}}[\mathrm{SiV}_{\mathrm{ground}}^{0}](L)\text{.}
\label{EqST12}
\end{equation}
Here, $E_{\mathrm{tot}}[\mathrm{SiV}{}_{\mathrm{excited}}^{0}](L)$ is the total energy of the \siv{} center in its $|^3A_{2g}\rangle$ ground state in a supercell with lattice constant $L$. The second term $E_{\mathrm{tot}}[\mathrm{SiV}_{\mathrm{ground}}^{0}](L)$ depicts the total energy of the excited state. We showed in a previous study~\cite{Thiering2019} that the product Jahn-Teller effect plays a significant role in the excitation process of \siv. Thus, we correct the pure electronic \textit{ab initio} data by the following 29.7~meV correction,
\begin{equation}
\Delta E_{\mathrm{loc}}^{\mathrm{corrected}}(L)=\Delta E_{\mathrm{loc}}^{\mathrm{HSE06}}(L)+\underset{=-29.7\;\mathrm{meV}}{\underbrace{\Delta E_{\mathrm{loc}}^{\mathrm{polaronic}}(L=14.26 \mathrm{\AA})-\Delta E_{\mathrm{loc}}^{\mathrm{HSE06}}(L=14.26 \mathrm{\AA})}}\text{,}
\label{EqST13}
\end{equation}
where $\Delta E_{\mathrm{loc}}^{\mathrm{polaronic}}(L=14.26 \mathrm{\AA})=1.344$~eV is the product Jahn-Teller ground state in the 512-atom diamond supercell and $\Delta E_{\mathrm{loc}}^{\mathrm{HSE06}}(L=14.26 \mathrm{\AA})=1.374$~eV is the value as obtained from Eq.~\eqref{EqST12}. Here, we assume that the correction from the product Jahn-Teller effect does not depend on the size of the supercell.

\subsubsection{$n=\infty$ transition, charge transition level}

Determining correct charge transition levels of defects always has been a difficult task in the supercell method. The origin of the inaccuracy is the Coulomb interaction which converges to zero with a long range $1/L$ strength. In a supercell that embeds a charged point defect, the mirror images of the charged defect and the compensating jellium charge in the plane wave supercell model interact via the long range Coulomb interaction. There are various correction schemes developed over the years to correct this artificial interaction such as Makov and Payne~\cite{Makov1995} (MP), Freysoldt, Neugebauer, and Van de Walle~\cite{Freysoldt2009} (FNV), and Lany and Zunger~\cite{Lany2008} (LZ). However, while the correction schemes provide accurate charge transition levels for supercells for a given size for moderately localized defect states, the most reliable method is to calculate the charge transition level with various supercells and fit the $1/L$ strength of the Coulomb interaction. Thus we determined the charge transition energy as follows
\begin{equation}
\Delta E_{n\rightarrow\infty}^{\mathrm{HSE06/PBE}}(L)=E_{\mathrm{tot}}[\mathrm{SiV}{}_{\mathrm{ground}}^{-}](L)-E_{\mathrm{tot}}[\mathrm{SiV}_{\mathrm{ground}}^{0}](L)-\varepsilon_{\mathrm{VBM}}(L)\text{,}
\label{EqST14}
\end{equation}
which consists of three individual DFT calculations. The first $E_{\mathrm{tot}}[\mathrm{SiV}{}_{\mathrm{ground}}^{-}](L)$ term is the total energy of the \sivm{} defect. The second $E_{\mathrm{tot}}[\mathrm{SiV}_{\mathrm{ground}}^{0}](L)$ is the total energy of the \siv{} defect. The third $\varepsilon_{\mathrm{VBM}}(L)$ is the valence band maximum of the perfect supercell. This is the Kohn-Sham eigenvalue of the highest occupied band of the diamond supercell with $L$ lattice size, where the Brillouin zone is sampled only at the $\Gamma$-point. We fit the following curve to our data to approach the $L\rightarrow+\infty$ bulk limit~\cite{Komsa2012}:
\begin{equation}
\Delta E_{n\rightarrow\infty}^{\mathrm{HSE06/PBE}}(L)=\frac{A}{L}+\frac{B}{L^{3}}+C\qquad\qquad\Delta E_{n\rightarrow\infty}^{\mathrm{HSE06/PBE}}(L\rightarrow\infty)=C\text{.}
\label{EqST15}
\end{equation}
The $\frac{A}{L}$ term is the long range monopole term of Coulomb interaction, the second $\frac{B}{L^{3}}$ is the quadrupole term, while we seek the value of $C$ corresponding to the isolated defect limit. We fit this formula to our \textit{ab initio} data by means of HSE06 and PBE DFT functionals. The results are depicted in Fig.~\ref{FigST6}(a). HSE06 provides 1.55~eV which is in excellent agreement with the experimental data (1.53~eV).

\subsubsection{Screening Effects in the $n=1$ Transition}

The $n=1$ localized excitation can be considered as a two-particle system. The first particle in \sivm{} traps a positively charged hole. We study an analogous system with a hydrogen atom confined into a small supercell. There the positively charged proton attracts the negative electron. When viewed from a remote distance, both systems are localized and neutral. If the supercell size is much larger than the Bohr radius of the hydrogen atom-like system ($a_0$) or $r_{n=1}$ radius for SiV defect (see Table~\ref{TabST1}), then the effect of the charged central particle is screened. This screening effect can have non-trivial effects on the functional form of the calculated energy as the supercell size is varied.

First, we determine the ratio of the screening length ($D_\mathrm{H}$) versus the  Bohr radius of hydrogen ($a_0$). Then, we use this ratio to approximate the screening length for the $n=1$ bound exciton excitation. We note that this screening effect does not happen for the $n=+\infty$ case as the VBM electron is delocalized all over the diamond lattice.

It can be observed in Fig.~3(b) that the $n=1$ state is poorly described by the simple model. The origin of this correction is that the $n=1$ orbital is expelled from the central region of the defect as the $e_g$ orbital of \sivm{} already occupies this region. Therefore, the binding energy will be increased by $\Delta = 0.23$~eV, and the effective radius ($r_\mathrm{eff}$) will be larger than that of the simple Bohr model ($r_{n=1}$). However we also take into account that the 1s orbital is active by an \textit{ungerade} phonon. Therefore the real central cell correction is only $\Delta - \hbar\omega_{A_{2u}}$. Here we determine this effective radius for the $n=1$ case. We search for the effective $n_\mathrm{eff}$ that reproduces the  excitation-energy ($E_\mathrm{eff}$) by means of central cell correction as given by
\begin{equation}
E_\mathrm{eff}=E_{I}-E_{y}/n_\mathrm{eff}^{2}=\underset{=E_{n=1}}{\underbrace{E_{I}-E_{y}}}+\Delta - \hbar\omega_{A_{2u}} \quad\Rightarrow\quad n_\mathrm{eff}=\sqrt{\frac{E_{y}}{E_{y}-(\Delta-\hbar\omega_{A_{2u}})}}=\sqrt{\frac{0.4}{0.4-0.23+0.0434}}=1.37\text{,}
\label{EqST18}
\end{equation}
where $E_I$ is the ionization energy.
Thus the effective Bohr radius is more than twice as large as the simple hydrogen model would indicate for the $n=1$ state,
\begin{equation}
r_\mathrm{eff}=r_{0}n_\mathrm{eff}^{2}=1.87\cdot r_{0}\text{.}
\label{EqST19}
\end{equation}

\begin{figure}[h!]
\includegraphics[scale=0.64]{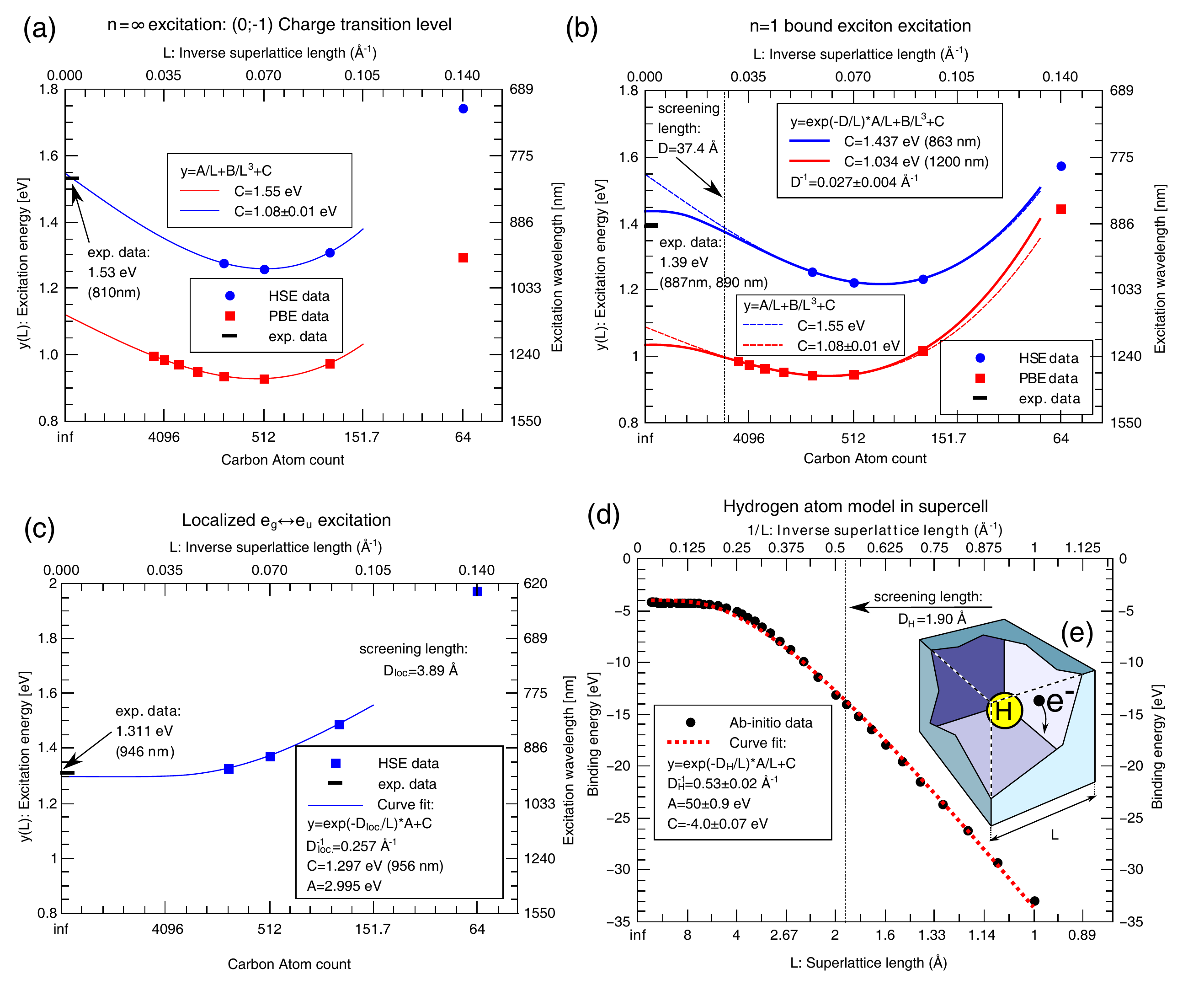}\caption{\label{FigST6} {\bf Excitation processes of \siv.} (a) Charge transition level of SiV between neutral and negative charge states by means of HSE06 and PBE functionals. We conclude that in the $L\rightarrow+\infty$ limit, our HSE06 results (1.55~eV) agree with the experimental data from Fig.~1 in the main text (1.53~eV). (b) $n=1$ bound exciton excitation by means of HSE06 and PBE functionals. Here we can see that the HSE06 limit at $L\rightarrow+\infty$ with screening included can explain the experimentally observed values at 1.39~eV. We note that the exponential part $\exp({-D/L})$ in the fit function relies on minor deviation in the meV regime, thus our result of $D^{-1} = 0.027 \pm 0.004 \mathrm{\AA^{-1}}$ is unambiguous. (c) Scaling of the $e_g\leftrightarrow e_u$ excitation process by means of HSE06 functional. (d) Hydrogen atom model in a $\Gamma$-point only calculation in a cubic supercell. (e) Schematic of the hydrogen atom in vacuum. The electron is effectively closed into a $L^3$ box. However, it is effectively not a box, as its warps around its edges. Thus the H atom bonds with itself. From a sufficiently large $L \gtrsim 4 \mathrm{\AA}$ distance, the H-atom in the supercell can be interpreted as a free non-interacting H-atom. We note that all calculations shown here are calculated only at the $\Gamma$-point.
}
\end{figure}

We also determine the ratio of the screening length ($D$) to the Bohr radius ($r_0$). We model this with a hydrogen atom in the simple cubic supercell with lattice constants ($L$). Since this system contains a single electron we apply the exact Hartree-Fock method in the calculation with a soft PAW potential with a plane wave cutoff of 200~eV. The results are shown in Fig.~\ref{FigST6}(d). One can clearly see that the Coulombic scaling ($-1/L$) deviates at sufficiently large supercells. We note that this scaling behavior is independent of the choice of the ionic potential of the proton (not shown) and the calculated total energies have a constant shift in each supercell. From a sufficiently far vantage point, the hydrogen atom is a neutral object, thus at large supercells ($L>4$~\AA), the energy of the system converges exponentially to a constant energy. We fit the total energies by
\begin{equation}
E_{H}(L)=\frac{A}{L}\exp\left(-\frac{D_{H}}{L}\right)+C\text{,}
\label{EqST20}
\end{equation}
where the Coulomb interaction ($A/L$) is screened by $\exp\left(-\frac{D_\mathrm{H}}{L}\right)$. We note that the repulsive $1/L^3$ term is missing because only a single proton appears in the system. Using this fitting procedure, we find $D_\mathrm{H} = 1.90 \,\mathrm{\AA} = 3.56 \cdot a_0$. The screening radius for the periodic array of hydrogen atoms is 3.56 times larger than the Bohr radius ($a_0=0.53 \mathrm{\AA}$ $n=1$) of the isolated free hydrogen atom.

Combining these results gives
\begin{equation}
D=\frac{D_\mathrm{H}}{a_{0}}r_{\text{eff}}=\frac{D_\mathrm{H}}{a_{0}}r_{0}n_{\text{eff}}^{2}=\frac{D_\mathrm{H}}{a_{0}}\frac{m_{e}}{m^{\star}}\frac{\varepsilon}{\varepsilon_{0}}a_{0}n_{\text{eff}}^{2}=\frac{m_{e}}{m^{\star}}\frac{\varepsilon}{\varepsilon_{0}}n_{\text{eff}}^{2}D_\mathrm{H}\text{.}
\label{EqST21}
\end{equation}

As a consequence, the effective screening lengths based on theoretical considerations and the hydrogen atom model for heavy hole, split-off hole, light hole, respectively, are the following:
\begin{subequations}
\begin{equation}
  D_{\mathrm{heavy\:hole}}=9.6\:\mathrm{\AA}
\end{equation}    
\begin{equation}
  D_{\mathrm{split-off\:hole}}=19.2\:\mathrm{\AA}
\end{equation}
\begin{equation}
  D_{\mathrm{light\:hole}}=29.0\:\mathrm{\AA} \text{,}
\end{equation}
\end{subequations}
which approaches $D=37.4\:\mathrm{\AA}$ as obtained from the fit to \textit{ab initio} data. Therefore, we can rationalize and explain the existence and magnitude of the screening length. This highlights again that the hydrogen-model for the $n=1$ bound exciton excited state is a poor approximation. The accuracy of this model is greatly improved for $n>1$ excited states. 

\subsubsection{$n=1$ Bound Exciton Transition}

We determine the excitation energy to the lowest $n=1$ bound exciton state with the following formula,
\begin{equation}
\Delta E_{n=1}^{\mathrm{HSE06/PBE}}(L)=E_{\mathrm{tot}}[\mathrm{SiV}{}_{n=1}^{BE}](L)-E_{\mathrm{tot}}[\mathrm{SiV}_{\mathrm{ground}}^{0}](L)\text{,}
\label{EqST16}
\end{equation}
which is very similar to Eq.~\eqref{EqST12} except that we have replaced the excited state with $E_{\mathrm{tot}}[\mathrm{SiV}{}_{n=1}^{BE}](L)$. Upon inspection of the $n=1$ and $n=+\infty$ results, both cases exhibits the long range Coulomb $\frac{A}{L}+\frac{B}{L^{3}}+C$ scaling. However, the analysis in the previous section shows that the Coulomb interaction is screened for the $n=1$ state. From a sufficiently large distance, the \sivm{} center plus a bound hole is a neutral object. Therefore at $L\rightarrow+\infty$ the $1/L$ scaling should be overtaken by a fast converging function such as screening damped by an exponential function. Thus we used the following fit function for the bound exciton resonances,
\begin{equation}
\Delta E_{n=1}^{\mathrm{HSE06/PBE}}(L)=\frac{A}{L}\exp\left(-\frac{L}{D}\right)+\frac{B}{L^{3}}+C\qquad\qquad\Delta E_{n\rightarrow1}^{\mathrm{HSE06/PBE}}(L\rightarrow\infty)=C\text{.}
\label{EqST17}
\end{equation}
We determine the screening length from PBE results ($D^{-1}=37.4\pm 5.1 \mathrm{\AA}$), where $A, B, C, D$ are fit parameters. Next, we use the fitted D to constrain HSE06 results, thus we assume that the $D$ is fixed, and allowed the remaining $A, B, C$ parameters to fit our data. Additionally we need to take into account that the $n=1$ state is optically not allowed by itself since it can be only observed by coupling to the $A_{2u}$ localized vibration mode of the Si atom. Therefore one has to add the frequency (energy) of this oscillatory motion $\hbar\omega_{A_{2u}} = 43.4$~meV (see Table~I in Ref.~\cite{Londero2018}) to the excitation energy which results in
\begin{equation}
\Delta E_{n\rightarrow1}^{\mathrm{corrected}}(L)=\Delta E_{n=1}^{\mathrm{HSE06}}(L)+\hbar\omega_{A_{2u}}\text{.}
\label{EqST18a}
\end{equation}

\subsection{Central Cell Correction and $n=2$ Transitions}
\label{Sec:n2RydbergDFT}
\begin{figure}[h!]
\includegraphics[width=0.9\columnwidth]{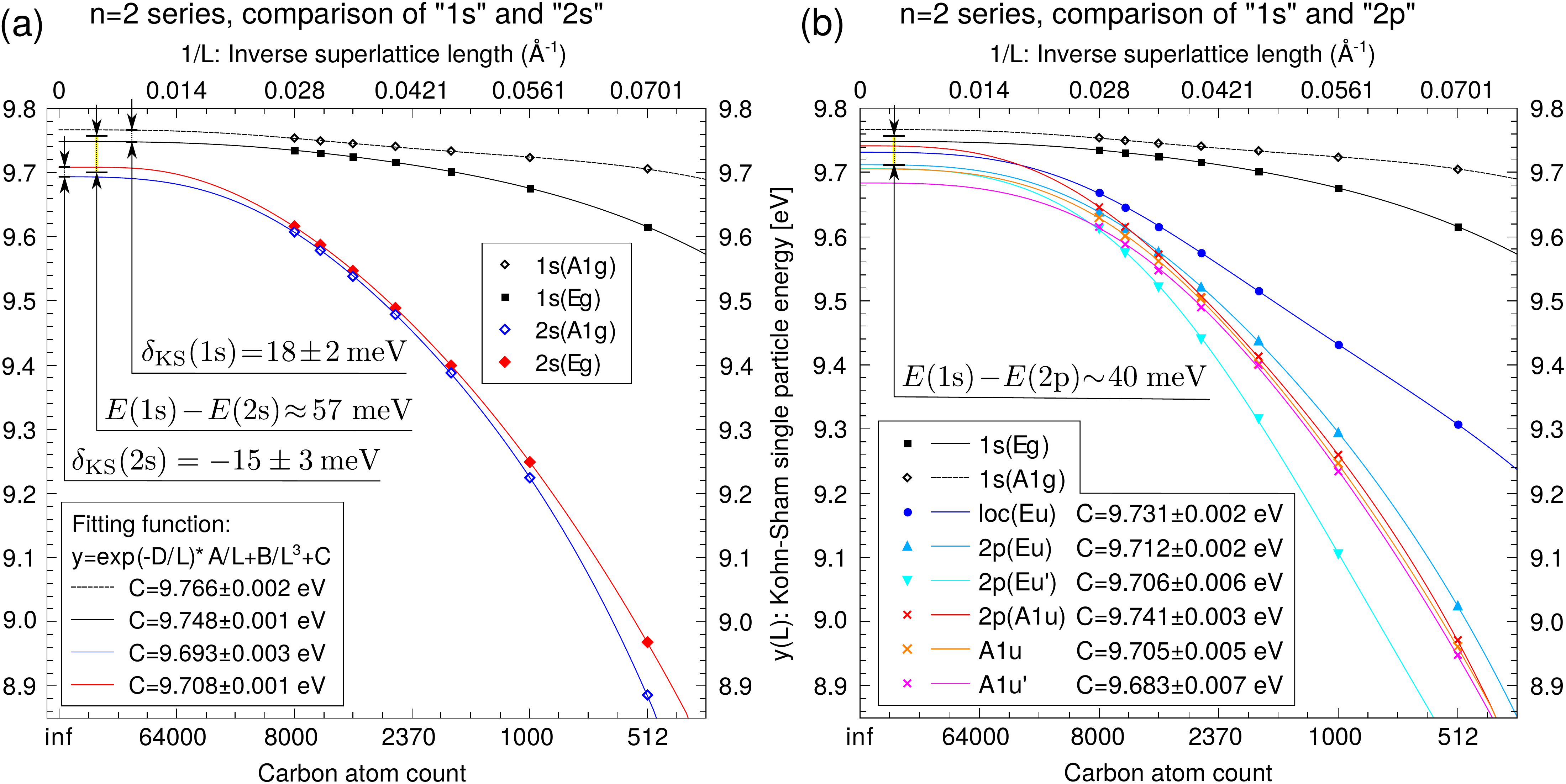}\caption{\label{FigST7}{{\bf Kohn-Sham levels of $n=2$ bound exciton states.} We note the fitting of $n=1$ orbitals crystal-field splitting parameter $\delta_\mathrm{KS}(1s)$ are not as good as in Fig.~\ref{FigST3}(a) because we fit the $1s(A_{1g})$ and $1s(E_{g})$ single particle energies individually (and not their difference) with a slightly different fit function. (a) ``2s'' bound exciton resonances are below the ``1s'' orbitals by $\approx57$~meV. (b) ``2p'' bound exciton resonances are below the ``1s'' orbitals by $\sim40$~meV. We note that the fit on the data of ``2p'' levels is very tentative. However, the sign of ``2s'' crystal-field splitting is seemingly the opposite to that of the $1s$ ($\delta_\mathrm{KS}(2s)$). We note that the energy difference of 2s and 2p can not directly compared with experiments ($E(\mathrm{2p})-E(\mathrm{2s})=17\:\mathrm{meV}$). We have seen in Section \ref{Sec:Isotopes} that the ``s'' orbitals are only active with an assisting \textit{ungerade} phonon, thus we need to add (see Eq.~\eqref{EqST18a}) the energy of that phonon (see Table~I in Ref.~\onlinecite{Londero2018}) to the ``2s'' orbital. Thus the calculated transition energy for the $2s$ state is 60~meV ($17+43.4=60.4\:\mathrm{meV}$.) \textit{higher} than the $2p$ state by means of our DFT results. }}
\end{figure}

We could also tentatively converge some of the $n=2$ resonances in the DFT calculations. We are unable to calculate these states by means of the accurate, but computationally-demanding, $\Delta$SCF calculations. Nevertheless, Kohn-Sham orbitals and levels of \sivm{} provide additional information beyond the ``1s'' states. We plot energies (not shown in Table~\ref{TS4}) below the $e_u$ localized orbital in Fig.~\ref{FigST7}. The ``2s'' excitation splits into $A_{1g}\oplus E_{g}$ in a similar manner as the ``1s''. The energy separation between them is approximately $E(\mathrm{1s})-E(\mathrm{2s})\approx57$~meV, much larger than the separation between the ``2p'' and ``1s'' series, $E(\mathrm{1s})-E(\mathrm{2p})\approx40$~meV. We note that the ``2s'' orbitals do not mix with the ${}^3E_u$ state because of the different parity of the wavefunctions but the ``2p'' orbitals can in principle mix with the ${}^3E_u$ state, making an accurate calculation demanding. 

There are three times as many ``2p'' than ``2s'' states, $\overset{\mathrm{VBM}}{\overbrace{(A_{1g}\oplus E_{g})}}\otimes\overset{\mathrm{2p}}{\overbrace{(A_{2u}\oplus E_{u})}}=3E_{u}\oplus2A_{2u}\oplus2A_{1u}$, because there are three 2p orbitals ($\mathrm{p_x}$, $\mathrm{p_y}$, $\mathrm{p_z}$), in addition to three orbitals split from the VBM states. According to \textit{ab initio} calculations, we can assign a set of $e_u$ and $a_{1u}$ orbitals for the $n=2$ series in Fig.~\ref{FigST7}(b) but cannot find the two $a_{2u}$ orbitals. We suspect that the 8000-atom supercell is still too small to accommodate these states, thus our predictions about ``2p'' states can only be tentatively compared to the experimental data. 

Nevertheless, our results clearly indicate that the ``1s'' is followed by the ``2p'' and not by the ``2s''. This is not a surprising result in the light of the fact that the ``1s'', ``2s'' orbitals are totally symmetric and have maximum density of their wavefunctions at the core of the \sivm{} defect, making them subject to a more significant central-cell correction.

If we compare this result against the experimental spin polarization and optical absorption spectra then one can tentatively assign the spin-polarization peaks, which are not optically active at $\sim$837 nm [see Fig.~3(a)], to the ``2s'' bound exciton resonance. Finally, we note that we cannot fully exclude the possibility that ``3d'' orbitals can also play a role in the optical spin polarization with photo-excitation near 830-840~nm wavelength region in the spin polarization spectra.

\clearpage

\end{document}